\documentclass[pra,aps,superscriptaddress,twocolumn]{revtex4}

\usepackage{graphicx}
\usepackage{amsfonts}
\usepackage{amssymb}
\usepackage{amsmath}
\usepackage{bm}
\usepackage{textgreek}
\usepackage{color}
\usepackage{bbm}
\usepackage{hyperref}

\newcommand{\Eq}[1]{Eq.\,(\ref{#1})}
\newcommand{\Eqs}[2]{Eqs.\,(\ref{#1})-(\ref{#2})}
\newcommand{\Fig}[1]{Fig.\,\ref{#1}}
\newcommand{\Figs}[2]{Figs.\,\ref{#1}-\ref{#2}}

\renewcommand{\r}{\mathbf{r}}

\newcommand{\Comp}{\mathcal{C}}

\newcommand{\I}{\mathbf{I}}
\newcommand{\C}{\mathbf{C}}
\renewcommand{\S}{\mathbf{S}}
\newcommand{\E}{\mathbf{E}}

\newcommand{\A}{\mathbf{A}}
\newcommand{\B}{\mathbf{B}}
\newcommand{\X}{\mathbf{X}}
\renewcommand{\d}{\pmb{\delta}}
\newcommand{\T}{^\mathrm{T}}
\newcommand{\N}{\mathbb{N}}

\newcommand{\x}{\mathbf{x}}

\renewcommand{\b}{\mathbf{b}}
\newcommand{\Ps}{\mathbb{P}}
\newcommand{\As}{\mathbb{A}}

\newcommand{\Capttwo}[1]{\caption{As Fig.\,S5, but for $\Comp_{#1}$.}}

\begin{document}

\title{Efficient quantitative hyperspectral image unmixing method for large-scale Raman micro-spectroscopy data analysis}
\author{E.\,G.~Lobanova}\email{Moreva@physics.msu.ru}\affiliation{School of Biosciences, Cardiff University, Cardiff CF10 3AX, United Kingdom}
\author{S.\,V.~Lobanov}\email{Lobanov@physics.msu.ru}\affiliation{School of Physics and Astronomy, Cardiff University, Cardiff CF24 3AA, United Kingdom}

\begin{abstract}
Vibrational micro-spectroscopy is a powerful optical tool, providing a non-invasive label-free chemically specific imaging for many chemical and biomedical applications. However, hyperspectral image produced by Raman micro-spectroscopy typically consists of thousands discrete pixel points, each having individual Raman spectrum at thousand wavenumbers, and therefore requires appropriate image unmixing computational methods to retrieve non-negative spatial concentration and corresponding non-negative spectra of the image biochemical constituents. 
In this article, we present a new efficient Quantitative Hyperspectral Image Unmixing (Q-HIU) method for large-scale Raman micro-spectroscopy data analysis.
This method enables to simultaneously analyse multi-set Raman hyperspectral images in three steps: (i) Singular Value Decomposition with innovative Automatic Divisive Correlation (SVD-ADC) which autonomously filters spatially and spectrally uncorrelated noise from data; (ii) a robust subtraction of fluorescent background from the data using a newly developed algorithm called Bottom Gaussian Fitting~(BGF); (iii) an efficient Quantitative Unsupervised/Partially Supervised Non-negative Matrix Factorization method~(Q-US/PS-NMF) via a fast combinatorial alternating non-negativity-constrained least squares algorithm, which rigorously retrieves non-negative spatial concentration maps and spectral profiles of the samples' biochemical constituents with no \textit{a priori} information and with great operation speed. Alternatively, the Q-US/PS-NMF is capable to work as a partially supervised method, when one or several samples' constituents are known, which significantly widens chemical and biomedical applications of the method. We apply the Q-HIU to the analysis of real large-scale Raman hyperspectral images ($\simeq${}$10^6$-by-$10^3$ combined data matrix) of human atherosclerotic aortic tissues and our results show a proof-of-principle for the proposed method to retrieve the biochemical composition of the tissues.
\end{abstract}

\maketitle

\section{Introduction}\label{sec:intro}

Over the past few decades, the development in multivariate image analysis methods has paved the way to handle and interpret the biochemical information, received from spectroscopic images quite efficiently.
For instance, spontaneous Raman micro-spectroscopy data have been treated using multivariate image reconstruction methods~\citep{Miljkovic2010,Krafft2017}, such as Principal Component
Analysis~(PCA), Hierarchical Cluster Analysis~(HCA), and K-Means cluster Analysis~(KMA), utilizing spectral contrast originated from biochemical composition variations over sample image pixels in order to produce pseudo-color images revealing this contrast. 
However, in these methods the spatio-spectral information is sorted into components that do not represent individual biochemical species with physically meaningful spectra and concentration.

Raman data have also been analysed using Vertex Component Analysis (VCA)~\citep{Hedegaard2011,Krafft2016,Czamara2017,Michael2017} and Multivariate Curve Resolution~(MCR)~\citep{Andrew1998}, which is also known as Non-negative Matrix Factorization (NMF). In contrast to the VCA which requires the presence of pixels containing pure biochemical substances, the MCR/NMF solves a general hyperspectral image unmixing problem without this restriction and therefore has broader chemical and biomedical applications.
Notably, in~Ref.~\citep{Andrew1998} MCR is based on Alternating Least Squares algorithm~(ALS), where non-negativity constrain is fulfilled by simply replacing all negative values with zeros. While this algorithm is straightforward, it normally converges to a wrong solution. 

In relation to Coherent Raman Scattering~(CRS) micro-spectroscopy~\citep{Camp2015,Cheng2015}, Coherent Anti-Stokes Raman scattering~(CARS) and Stimulated Raman scattering~(SRS) images have been analysed by PCA~\citep{Lin2011}, cluster analysis based on spectral phasor approach~\citep{Fu2014}, MCR~\citep{Zhang2013} and NMF~\citep{DiNapoli2016} both based on Alternating Non-negativity-constrained Least Squares algotithm~(ANLS).   
In contrast to PCA and HCA, MCR/NMF can provide a quantitative determination of the chemical composition.
Importantly, the popular MATLAB realization of the MCR-ANLS algorithm\,\citep{Jaumot2015} uses computationally inefficient realization of non-negative least squares (NLS) algorithm, which can be several orders of magnitude slower than fast combinatorial NLS (FC-NLS)\,\citep{VanBenthem2004,Kim2007a}.
Indeed, a standard NLS algotithm\,\citep{Lawson1995} is designed to calculate a non-negative right hand-side~(RHS) vector.
The multiple-RHS problem, used to quantify the chemical information in a hyperspectral image, can be reduced to the independent solution of problems for each RHS vector. However, the computation of this technique is time consuming, which becomes even worse for big data-sets.
For instance, computational speed of the MCR-ANLS implemented by \citep{Jaumot2015} is about 200 times slower than our realization of the FC-ANLS (see Fig. S1 of the Supplementary Material).
NMF via FC-NLS algorithm~(FC-NMF) benefits from good computational speeds, but a MATLAB realization of this method\,\citep{Kim2007a} is verified (see Fig. S2 of the Supplementary Material) to be about three times slower as compared with our realization, which can be significant for large volume data analysis.
Overall, fast and accurate numerical codes solving the NMF problem are still under intense scrutiny.

The goal of our paper is to develop efficient Quantitative Unsupervised/Partially~Supervised Hyperspectral Image Unmixing (Q-HIU) method for the analysis of large-scale Raman micro-spectroscopy data.

\section{Theory}
\subsection{Formulation of the hyperspectral image unmixing problem}
\paragraph{Forward scattering problem.}
Suppose the examining specimens are a mixture of $N$ pure chemicals (components) with known individual Raman scattering cross-section spectra $S_i(\nu)$, $i=1,2,\ldots,N$ and known spatial distributions of concentration $C_i(\r)$.
Here, $\nu$ indicates Raman shift and $\r=(x,y;r)$ is a radius vector drawn to the image pixel with coordinates $(x,y)$ of the $r^\mathrm{th}$ hyperspectral image.
For this example problem, one could measure Raman intensity matrix $I(\r;\nu)$, which is a linear combination of the individual spatially resolved concentrations and corresponding pure spectra products of the samples' constituents

\begin{equation}
I(\r;\nu) = \sum_{i=1}^N C_i(\r)S_i(\nu).\label{Eq:IFactorization}
\end{equation}
This equation represents the solution of the forward scattering problem, which is computationally straightforward and not interesting.

\paragraph{Inverse scattering problem.}
A more challenging problem, which is usually the case of chemical and biomedical applications, is to retrieve the chemical composition of samples without prior knowledge of their constituents.
This inverse scattering problem can be formulated in a following way.
For a given Raman scattering cross-section intensity $I(\r;\nu)$, one should factorize it into $N$ separate chemical components with individual Raman scattering cross-section spectra $S_i(\nu)$ and spatially resolved concentrations $C_i(\r)$.
In this paper, we also consider the case, where one or several Raman spectra $S_i(\nu)$ can be known.

In spontaneous point-scan Raman experiment, hyperspectral image (see illustration of a hyperspectral data cube in \Fig{Fig2}A) is acquired by raster scanning the sample through the focal point of the laser beam, and therefore does not consist of the continuously varying coordinates $x$, $y$ and wavenumber $\nu$, but  finite sets of discrete points: $x=\{x_i\colon \, i = 1, 2, \ldots, N_x(r)\}$, $y=\{y_i\colon \, i = 1, 2, \ldots, N_y(r)\}$, $\nu=\{\nu_i\colon \, i = 1, 2, \ldots, N_s\}$.
Here, $N_x(r)$ and $N_y(r)$ are a number of pixel points in the $r^\mathrm{th}$ hyperspectral image along $x$- and $y$-axes, respectively, and $N_s$ is a number of spectral points of this image.
We can also construct a set of radius vectors $\r=\{\r_i\colon \, i = 1, 2, \ldots, N_p\}$, where $N_p=\sum_{r=1}^{N_r} N_x(r)N_y(r)$ is a total number of pixel points and $N_r$ is a total number of hyperspectral images.
This enumeration allows us to represent the Raman intensity $I(\r;\nu)$, spatial concentrations $C_i(\r)$ and spectra $S_i(\nu)$ as matrices $I_{ij}=I(\r_i;\nu_j)$, $C_{ij}=C_i(\r_j)$, $S_{ij}=S_i(\nu_j)$, and express \Eq{Eq:IFactorization} in the matrix form
\begin{equation}
\I = \C\T\S, \label{Eq:IFactMatrix}
\end{equation}
where the superscript $\T$ means matrix transpose.
In what follows we will use both matrix and explicit forms of the Raman intensity, concentration, and spectra.

\subsection{Singular Value Decomposition with Automatic Divisive Correlation~(SVD-ADC) for noise filtering}\label{Sec:Data1}
A standard formulation of SVD algorithm \citep{Golub2013} applied to hyperspectral intensity matrix $\I$ enables to project the data in a new spatio-spectral orthonormal basis consisting of pseudo-concentrations $C_i(\r)$ and pseudo-spectra $S_i(\nu)$ 
\begin{equation}
I(\r;\nu) = \sum_{i\in \N} C_i(\r) \lambda_i S_i(\nu),\label{Eq:SVD}
\end{equation}
which can be sorted by their descending singular values $\lambda_i$~\citep{Uzunbajakava2003}.
Here, $\N=\{1,2,\ldots,\mathrm{rank}(\I)\}$ and $C_i(\r)\cdot C_j(\r)=S_i(\nu)\cdot S_j(\nu)=\delta_{ij}$, where the dots indicate Euclidean scalar products (see Supplementary Material for definition).

It is believed that physically meaningful components have significantly higher singular values compared to noise ones. However, an adequate threshold criterion, which rejects relatively small singular values and its corresponding singular vectors, to the best of our knowledge, does not exist in literature.
To overcome this limitation, we propose a SVD in a new formulation, which we call SVD with Automatic Divisive Correlation~(SVD-ADC). This approach requires additional step once SVD is performed, i.e.: for each left and right singular vectors ($C_i(\r)$ and $S_i(\nu)$, respectively) we calculate spectral $R^S_i$ and spatial $R^C_i$ autocorrelation coefficients at "one" pixel shift
\begin{gather}
R^S_i = R\left[ S_i(\nu), S_i^\delta(\nu) \right], \\
R^C_i = \max_{\d} R\left[ C_i(\r), C_i^{\d}(\r) \right],
\end{gather}
where $R[\ldots]$ denotes correlation coefficient (see Supplementary Material for definition),
$S_i^\delta(\nu)$ indicates pseudo-spectrum shifted by "one" spectral point, i.e. $S_i^\delta(\nu_j)=S_i(\nu_{j+\delta})$,
and $C_i^{\d}(\r)$ represents pseudo-concentration shifted by "one" image pixel along $x$- [$\d=(\delta_x,0)$] or $y$-axis [$\d=(0,\delta_y)$].
Here, double quotation marks for the word "one" implies that one pixel shift should be chosen only when the spectral/spatial resolution of the used instrumentation is smaller than the pixel size of a CCD camera.
For example, in our analysis presented below we used $\delta=\delta_x=3$ and $\delta_y=1$.

This approach enables to filter the meaningful components from noise-dominated ones (e.g.: shot noise, read noise, residual noise after cosmic rays removal, and etc.)  by plotting the spatial autocorrelation coefficients $R^C_i$ against spectral ones $R^S_i$ for each pair of singular vectors and discard  components with mean autocorrelation coefficient $R_i=(R^C_i+R^S_i)/2$ lower than $R_\mathrm{thr}=50$\%. This number was testified to adequately differentiate the meaningful components from the noise-dominated ones, which can be simply explained by the fact that an ideally noisy component exhibits 0\,\% autocorrelation coefficient, whereas it is approximately 100\% for an ideally meaningful component.
After replacing the set $\N$ in \Eq{Eq:SVD} by the set $\N_\mathrm{thr}=\{i\colon R_i > R_\mathrm{thr}\}$ and performing summation over this set, we find Raman intensity matrix with reduced/removed noise.

\subsection{Bottom Gaussian Fitting (BGF) for background subtraction}\label{Sec:Data2}
\begin{figure}
\includegraphics[width=\linewidth]{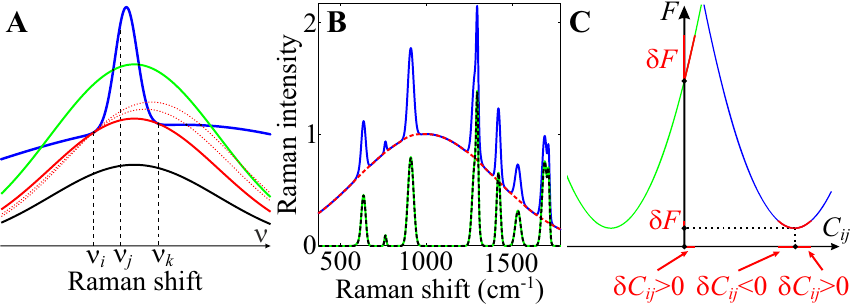}
\caption{
{\bf A}, Schematic representation of the background subtraction procedure using the BGF.
The input artificial spectrum resembling Raman-like band is shown by the blue line.
The black, red, and green lines show Gaussian functions with the same STDs $\sigma$, but different amplitudes $A$ and expected values $\mu$ (see text for details).
{\bf B}, Example of the BGF procedure showing efficiency of the developed algorithm.
The input artificial Raman-like spectrum $\I^\mathrm{a}=\I^\mathrm{aR}+\I^\mathrm{aB}$ is shown by the blue line.
The result of the BGF $\I^\mathrm{R}$ is shown by the green dashed line: the bottom Gaussian fit $\I^\mathrm{B}$ (red dashed line) is subtracted from the Raman-like spectrum $\I^\mathrm{a}$, resulting in the background-free Raman-like spectrum $\I^\mathrm{R}$, which is in a good agreement with the model $\I^\mathrm{aR}$ (black line, see text for details).
{\bf C}, Dependence of the factorization error $F(\C,\S)$ on a matrix element $C_{ij}$ for fixed other elements. Two special cases are shown: the blue parabola has a minimum at positive $C_{ij}$ value, whereas a minimum of the green parabola is in the forbidden region $C_{ij}<0$.
Red lines indicate variation of the factorization error $\delta F$ at the constrained minima: $C_{ij}>0$ and $C_{ij}=0$ for the first (blue parabola) and second (green parabola) cases, respectively.
}  
\label{Fig1}
\end{figure}
Raman spectra have an inherent broad background originating from fluorescence and amorphous scattering from glass substrate and sample itself, and therefore baseline correction algorithms via polynomial functions of different orders are commonly used for background removal\,\citep{Hirokawa1980,Baek2009}. We propose a new computational method, which we call Bottom Gaussian Fitting~(BGF), allowing to autonomously subtract a complex curved background from Raman micro-spectroscopy data resulting in the accurate quantification of the Raman bands in the spectra at each image pixel.
The principles of this mathematical approach are explained in the following.

For the sake of brevity, we will omit in this section the first index in the Raman intensity matrix $\I$ labelling image pixel, i.e. we will write $I_j$ instead of $I_{ij}$ meaning that the background removing procedure must be done independently for each image pixel $i$.

Let us split the Raman intensity $\I$ into two non-negative parts
\begin{equation}
\I = \I^\mathrm{R} + \I^\mathrm{B},
\end{equation}
where $\I^\mathrm{R}$ contains sharp resonance peaks representing Raman bands, whereas $\I^\mathrm{B}$ is a quasi-slowly varied background approaching the Raman intensity $\I$ from the bottom.
Our aim is to find $\I^\mathrm{B}$ as a quasi-superposition of Gaussian functions with standard deviations (STD) larger than $\sigma$ -- \emph{the only one parameter} that will be used for background subtraction.

Let us consider some Raman shift $\nu_j$ and find maximum possible background intensity at this point $I^\mathrm{B}_j$.
The figure \ref{Fig1}A illustrates a series of Gaussian functions (black, green, and red lines) with the fixed STD $\sigma$ and varying amplitude $A$ and expectation value $\mu$, which can potentially be background fits for a Raman spectrum $\I$ shown in \Fig{Fig1}A by the blue line.
As one can see from the figure, only the red solid line tangents the Raman spectrum $\I$, so that the representative Raman intensity $\I^\mathrm{R}$ at the point $\nu_j$ is well quantified.
Mathematically, since the background intensity $\I^\mathrm{B}$ tangents the Raman intensity $\I$, it should be determined as a Gaussian function $G_\sigma(\nu_j;\nu_i,I_i,\nu_k,I_k)$ with the STD $\sigma$, which passes through left-sided $(\nu_i,I_i)$ and right-sided $(\nu_k,I_k)$ points of the Raman intensity $\I$ relating to the considered Raman shift $\nu_j$.
Furthermore, the tangent background fit $\I^\mathrm{B}$ (red solid line) has minimum Raman intensity value at the point $\nu_j$ as can be visually observed from the \Fig{Fig1}A (compare the red solid line with the red dotted and green lines).
Thus, the bottom Gaussian fit can be defined as
\begin{equation}
I^\mathrm{B}_j = \min_{i \leq j<k}G_\sigma(\nu_j;\nu_i,I_i,\nu_k,I_k),\label{Eq:IBG}
\end{equation}
where the Gaussian function
\begin{equation}
G_\sigma(\nu_j;\nu_i,I_i,\nu_k,I_k) = A e^{-\frac{(\nu_j-\mu)^2}{2\sigma^2}}
\end{equation}
has amplitude $A$ and expected value $\mu$, which are implicitly  defined by the following equations
\begin{equation}
A e^{-\frac{(\nu_i-\mu)^2}{2\sigma^2}}=I_i, \quad
A e^{-\frac{(\nu_k-\mu)^2}{2\sigma^2}}=I_k.
\end{equation}

Computing \Eq{Eq:IBG} for all Raman shifts $\nu_j$, we find a quasi-slowly varying curve $\I^\mathrm{B}$ fitting the Raman intensity $\I$ from the bottom.
The difference between these two curves ($\I^\mathrm{R} = \I - \I^\mathrm{B}$) contains only sharp resonances or Raman bands.

To show the efficiency of the developed algorithm, we give an example of BGF procedure applied to an artificial spectrum displaying ten Raman bands~(\Fig{Fig1}B). This input spectrum was chosen to be a superposition of Gaussian functions resembling Raman-like signal $\I^\mathrm{aR}$ superimposed by a background $\I^\mathrm{aB}$. The background has unitary amplitude $A^\mathrm{aB}=1$, STD $\sigma^\mathrm{aB}=400$\,cm$^{-1}$, and expected value $\mu^\mathrm{aB}=1000$\,cm$^{-1}$. The amplitudes, STDs, and expected values for the ten Raman-like bands were randomly selected from the intervals $(0,1)$, $(5, 20)$\,cm$^{-1}$, and $(370, 1783)$\,cm$^{-1}$, respectively.
As one can see from \Fig{Fig1}B, the background-free Raman-like signal $\I^\mathrm{R}$ (green dashed line), resulted from background subtraction with the representative bottom Gaussian fit $\I^\mathrm{B}$ of $\sigma=300$\,cm$^{-1}$ (red dashed line), is in a good agreement with the Raman-like signal $\I^\mathrm{aR}$ (black line).

Once the background-free Raman-like signal $\I^\mathrm{R}$ is retrieved using BGF, we calculate the spatially-resolved spectral auto-correlation coefficients for this matrix at "one" pixel shift and then discard the spatial points, which have correlation values lower than 50\%, from the further analysis. This optional additional step allows to eliminate contribution of spatial points, which are strongly affected by noise due to dominated background (orders of magnitude larger than the remaining signal). 

\subsection{Efficient Quantitative Unsupervised/Partially Supervised Non-negative Matrix Factorization~(Q-US/PS-NMF)}\label{Sec:Dat3}
In a standard formulation, NMF approach \citep{Paatero1994,Paatero1997,Lee1999,Lee2001} allows to decompose the spatially-resolved Raman intensity matrix $\I$ into a product \Eq{Eq:IFactMatrix} of non-negative spatial concentration maps $\C$ and corresponding non-negative spectra $\S$ with rows representing individual biochemical substances~(components) of the samples' composition.
Both matrices $\C$ and $\S$ are unknown prior to NMF.

In this paper, we generalise NMF to operate as a partially supervised method, which we call Q-US/PS-NMF. This modality can be particularly useful when required to eliminate the contribution of wax residues to the spectral basis for paraffin-embedded samples. For this, the spectrum of paraffin-wax compound is fixed during the Q-US/PS-NMF procedure.

Mathematically, the problem can be formulated as follows.
Suppose $N_k$ biochemical substances with Raman spectra $S_i(\nu)$, $i=1,\ldots,N_k$ are known and our goal is to determine the rest $N_u=N-N_k$ spectra $S_i(\nu)$, $i=N_k+1,\ldots,N$, as well as concentration maps for both known and unknown components $C_i(\r)$, $i=1,2,\ldots,N$.
Denoting known/unknown parts of the spectral matrix $\S$ and corresponding parts of the concentration matrix $\C$ with letters $k$/$u$, we have
\begin{equation}
\C = \begin{pmatrix}
\C_k \\
\C_u
\end{pmatrix}, \quad
\S = \begin{pmatrix}
\S_k \\
\S_u
\end{pmatrix}.
\end{equation}
Ideally, these matrices should satisfy \Eq{Eq:IFactMatrix}.
However, for real experimental data, the Raman intensity matrix $\I$ can be factorized only approximately with a residue $\E$, i.e. $\I-\C^\mathrm{T}\S=\E$.
A pair of matrices $\C$ and $\S$, which minimises Frobenius norm of the residue $\|\E\|_\mathrm{F}$, represents the solution of the factorization problem \Eq{Eq:IFactMatrix}.
This is equivalent to the minimization of the following function
\begin{equation}
F(\C,\S)=\frac{1}{2}\|\I-\C^\mathrm{T}\S\|_\mathrm{F}^2\label{Eq:F}
\end{equation}
subject to $C_{ij}\geq 0$, $S_{ij}\geq 0$.
The necessary condition for a minimum of \Eq{Eq:F} is that a variation of the factorization error $\delta F$ at the point $(\C,\S)$ is non-negative, i.e. 
\begin{equation}
\delta F = \sum_{i=1}^{N}\sum_{j=1}^{N_p} \frac{\partial F}{\partial C_{ij}}\delta C_{ij} + \sum_{i=N_k+1}^{N}\sum_{j=1}^{N_s} \frac{\partial F}{\partial S_{ij}}\delta S_{ij}\geq 0.\label{Eq:dF}
\end{equation}
Note, since variations $\delta C_{ij}$ and $\delta S_{ij}$ are arbitrary, each term of this equation must be greater or equal to zero.

Let us consider dependence of the factorization error $F$ on a variable $C_{ij}$ for fixed other variables.
The graph of this function is parabola, which can have the minimum either at a point $C_{ij}>0$ (blue line in \Fig{Fig1}C) or in the forbidden region $C_{ij}<0$ (green line in \Fig{Fig1}C).
In the former, variation $\delta C_{ij}$ (highlighted by red color in \Fig{Fig1}C) can be both positive and negative, which requires the partial derivative $\partial F/\partial C_{ij}$ to be zero in order to satisfy \Eq{Eq:dF}.
In the latter, the constrained minimum must be shifted from the region with negative values of $C_{ij}$ to the nearest possible point in the non-negative region, which is $C_{ij}=0$, so the variation $\delta C_{ij}$ can have only positive values.
Thus, the partial derivative $\partial F/\partial C_{ij}$ is not required to be zero and can take positive values.
These requirements are called the Karush-Kuhn-Tucker conditions~\citep{Lawson1995} and can be written in a following way
\begin{eqnarray}
\frac{\partial F}{\partial C_{ij}} = 0 \; \mathrm{if} \; C_{ij} > 0, &\quad &
\frac{\partial F}{\partial C_{ij}} \geq 0 \; \mathrm{if} \; C_{ij} = 0, \label{Eq:dFdC}\\
\frac{\partial F}{\partial S_{ij}} = 0 \; \mathrm{if} \; S_{ij} > 0, &\quad &
\frac{\partial F}{\partial S_{ij}} \geq 0 \; \mathrm{if} \; S_{ij} = 0,\label{Eq:dFdS}
\end{eqnarray}
where the bottom formulas must be valid only for $i > N_k$, i.e. for unknown spectra.

Differentiating \Eq{Eq:F}, we find partial derivatives
\begin{gather}
\frac{\partial F}{\partial \C} = \S\S^\mathrm{T}\C - \S\I^\mathrm{T},\\
\frac{\partial F}{\partial \S_u} = \C_u\C_u^\mathrm{T}\S_u - \C_u\I + \C_u \C_k^\mathrm{T} \S_k,
\end{gather}
which lead to a non-linear problem after substitution them into \Eqs{Eq:dFdC}{Eq:dFdS}.
This non-linear problem can be solved iteratively by alternative fixation of one matrix ($\C$ or $\S_u$) and computation of another one ($\S_u$ or $\C$, respectively) using NLS algorithm~\citep{Lawson1995}.
Namely, alternately denoting $\S\S^\mathrm{T}$ and $\C_u\C_u^\mathrm{T}$ by $\A$, $\S\I^\mathrm{T}$ and $\C_u\I - \C_u \C_k^\mathrm{T} \S_k$ by $\B$, unknown non-negative matrices $\C$ and $\S_u$ by $\X$, and independently considering  each column $\x$ and $\b$ of the matrices $\X=(\x_1,\x_2,\ldots)$ and $\B=(\b_1,\b_2,\ldots)$, one find a solution of the NLS problem
\begin{equation}
\begin{aligned}
\x_p &= \A^{-1}_{pp} \b_{p}\\
\x_a &= 0
\end{aligned}
 \quad 
\mathrm{subject} \: \mathrm{to} \quad
\begin{aligned}
\x_{p} > 0, \\
\A_{ap}\x_{p}-\b_{a} \geq 0.
\end{aligned}\label{Eq:NLSsol}
\end{equation}
Here, the indexes $p$ and $a$ denote projection of the vectors $\x$, $\b$ and matrix $\A$ onto passive and active subspaces characterizing by two complementary sets $\Ps$ and $\As$, respectively:
\begin{equation}
\begin{array}{cc}
\Ps\cap \As = \varnothing, & \Ps\cup \As = \{1,2,\ldots\},\\
\x_p=\{x_i\colon i\in \Ps\}, & \x_a=\{x_i\colon i\in \As\},\\
\A_{pp}=\{A_{ij}\colon i\in \Ps, \, j\in \Ps\},\;  & \; 
\A_{ap}=\{A_{ij}\colon i\in \As, \, j\in \Ps\}
\end{array}\nonumber
\end{equation}
and similar for $\b$.

The described procedure can be done independently for each column of the matrix $\X$.
However, working with the whole matrices rather than vectors is more computationally efficient~\citep{VanBenthem2004,Kim2007a}.
Indeed, one can group the columns that share the same index sets $\Ps$ and compute the appropriate matrix inverse $\A^{-1}_{pp}$ in \Eq{Eq:NLSsol} for each block.
This procedure is apparently orders of magnitude faster than sequential computation of the matrix inverse for each individual column, which is used in the popular MATLAB realization of the MCR-ANLS algorithm by~\citep{Jaumot2015} (see Fig.\,S1 of the Supplementary Material).
Our realisation of NMF algorithm, which we call Q-US/PS-NMF, is implemented in a fast combinatorial ANLS framework and shows exponential convergence to a local minimum with three times faster computational performance compared to the popular MATLAB version of FC-NMF\,\citep{Kim2007a} (see Fig.\,S2 of the Supplementary Material).

Furthermore, the Q-US/PS-NMF algorithm allows to quantitatively determine and compare pixel-by-pixel concentrations (strictly speaking, Raman scattering cross sections) of different biochemical components between samples by applying the following normalization conditions during final step of Q-US/PS-NMF analysis: the spectrum of each factorized component is normalized on \textit{const} so that the mean spectral intensity for each component was the same and the sum of components' mean concentrations is set equal to one.
\begin{gather}
\int d\nu S_1(\nu) = \int d\nu S_2(\nu)=\ldots=\int d\nu S_N(\nu),\\
\frac{1}{N_p}\sum_{j=1}^{N_p} \sum_{i=1}^N C_i(\r_j) = 1.
\end{gather}

Overall, Q-HIU presented here enables to decompose large Raman multi-datasets, simultaneously analysed from a variety of samples, into individual chemical components with spatially resolved concentration maps and corresponding spectra, in a quantitative manner without prior knowledge of the chemical composition of samples or with partial knowledge of some constituents' spectra.

\section{Results and discussion}
\begin{figure*}
\includegraphics[width=\linewidth]{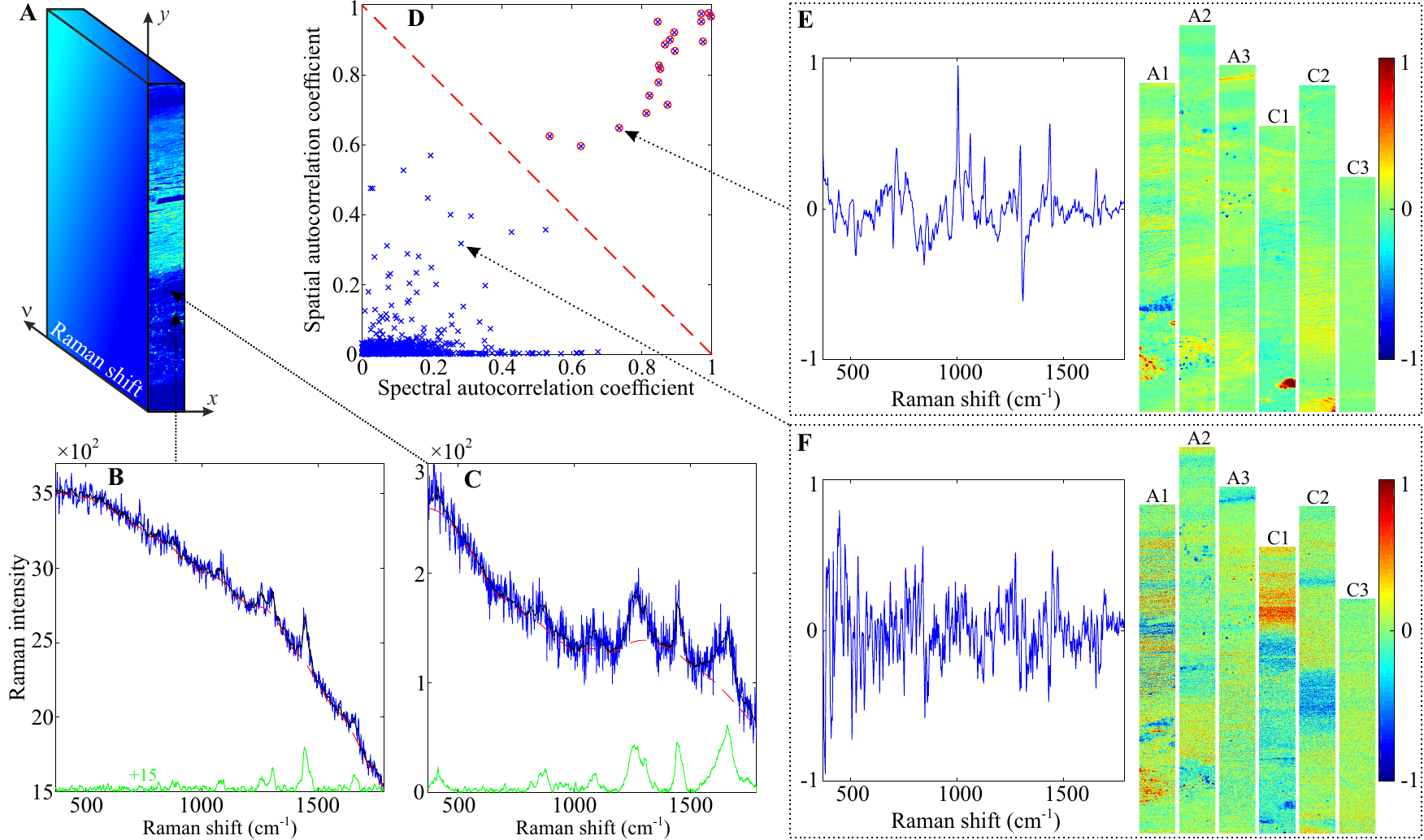}
\caption{{\bf A}, Schematic of a hyperspectral data cube. The arrows mark two pixels, Raman spectra of which are shown on the panels {\bf B} and {\bf C}.
{\bf B}-{\bf C}, The input Raman spectrum (blue line) \textit{before} the SVD-ADC and the corresponding noise-filtered Raman spectrum (black line) \textit{after} this procedure.
The results of background subtraction using the BGF is shown by the green line: the bottom Gaussian fit (red dashed line) is subtracted from the noise-filtered Raman spectrum, resulting in background-free Raman spectrum.
{\bf D}, Spatio-spectral autocorrelation coefficients map of singular vectors found from the SVD-ADC. The dashed diagonal line represents a decision line for mean autocorrelation coefficients $R_i$ at $R_\mathrm{thr}=\:$50\%, separating the coefficients above the line~(circled cross signs -- meaningful components) from those below it~(cross signs -- noise).
{\bf E}-{\bf F}, The spectra and spatial distributions of two singular vectors with the mean autocorrelation coefficients of (50$\pm$20)\% as indicated on the panel {\bf D}.
On both panels, 6 Raman images are labelled according to the sample source.
Scale bars are 100\,$\mu$m.
}  
\label{Fig2}
\end{figure*}
In the previous section, we presented a novel efficient Q-HIU approach for large-scale Raman micro-spectroscopy data analysis, consisting of three methods: SVD-ADC, BGF, Q-US/PS-NMF, each of which represents main step in analysis.
In the following, we demonstrate an example of the Q-HIU analysis applied to real experimental large-scale Raman micro-spectroscopy images of human thoracic aortic tissues, which were sourced from \citep{You2017b}.
The data consist of 3 large volume Raman images (see schematic of a hyperspectral data cube in \Fig{Fig2}A) of diseased aortic tissues, containing biochemical signatures of atherosclerotic plaque lesions (e.g.: foam cells/necrotic core, cholesterol crystals, calcification), and 3 Raman images of non-atherosclerotic age-matched control ones. In the following, we refer to atherosclerotic~(A) and non-atherosclerotic control~(C) samples as A1~(022$\_$p$\_$000), A2~(025$\_$p$\_$000), A3~(031$\_$p$\_$005), and C1~(043$\_$000), C2~(046$\_$000), C3~(047$\_$002), respectively (see~\citep{You2017} for details on tissue cohorts). The Q-HIU analysis was performed {\it simultaneously} on all Raman images, enabling to show the computational efficiency of the developed approach. The resulting hyperspectral Raman multi-image was a $\simeq$6$\times$10$^2$$\times$10$^3$$\times$10$^3$ matrix, where the first, second-third, and fourth dimensions represent the numbers of participating samples, spatial pixels in \textit{x}- and \textit{y}-directions, and spectral points, respectively.

The Raman data from biomedical samples to be analysed here are ideal candidates to show a proof-of-principle for accuracy of the proposed Q-HIU methodology to retrieve the biochemical composition of samples. 

Importantly, the new developed Q-US/PS-NMF method in a partially supervised formulation also presented in this paper is already used by us and co-authors in the analysis of large-scale label-free Raman hyperspectral images from biomedical paraffin-embedded samples in Alzheimer's disease\,\citep{Lobanova2018}. In that paper, wax component was fixed during the Q-US/PS-NMF procedure in order to avoid the spectral contamination of the biochemical components, characteristic for human brain tissue, by wax residues.

\begin{figure*}
\includegraphics[width=\textwidth]{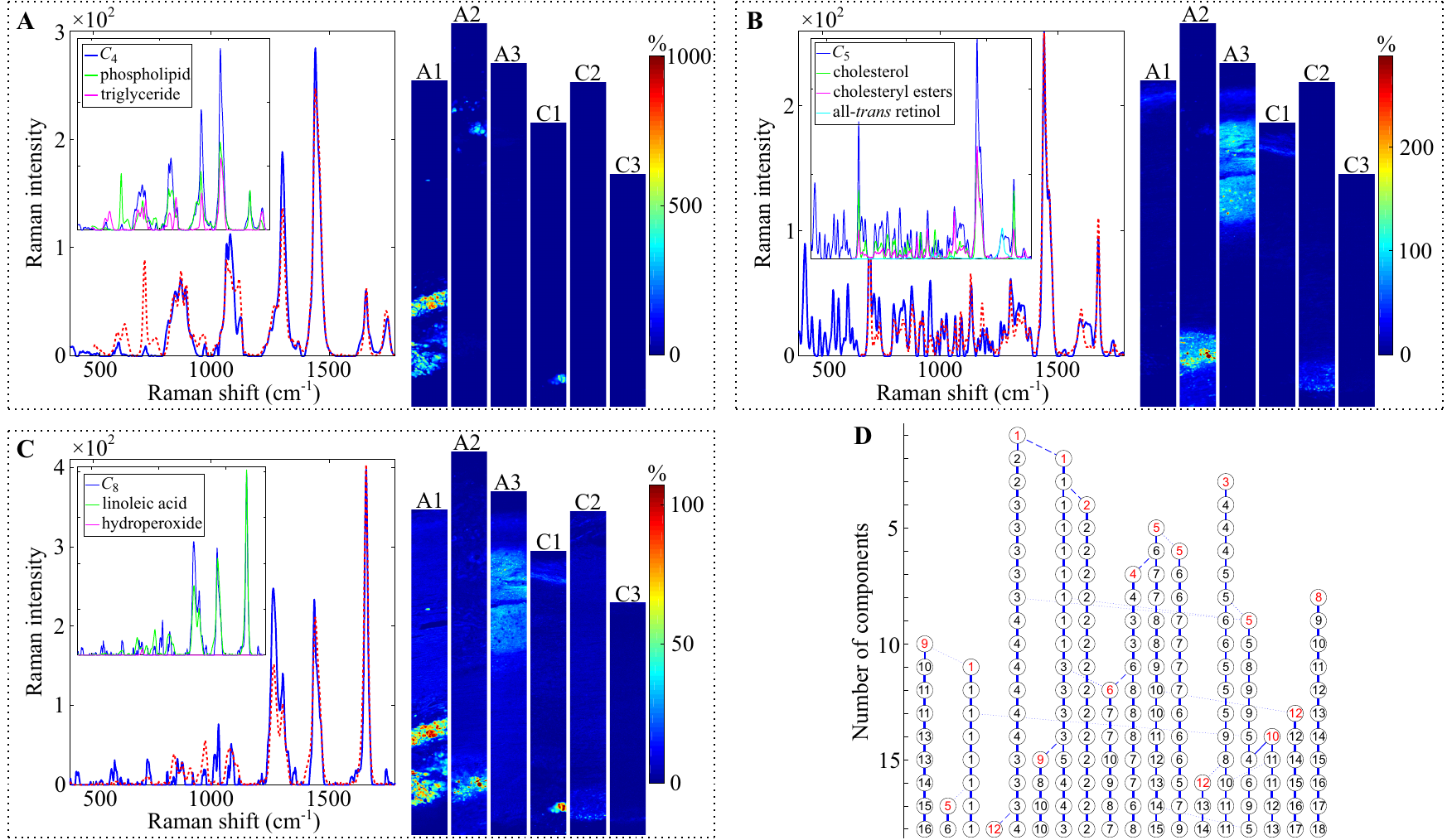}
\caption{{\bf A}-{\bf C}, Component spectra (blue lines) and spatial distributions of concentration of $\Comp_4$, $\Comp_5$, and $\Comp_8$ found from the Q-US/PS-NMF of 6 Raman images from atherosclerotic aortic tissue regions~(A1, A2, A3) and non-atherosclerotic controls~(C1, C2, C3). The chemical attribution of components
is based on the comparison with analytical standard Raman
spectra of pure chemical species. Fits are shown by red dashed
lines. Partial contributions (green, magenta, and black lines) to each component spectra (blue lines) are also indicated in the inset.
Scale bars are 100\,$\mu$m.
{\bf D},~Evolution of component spectra resulting from the Q-US/PS-NMF with the numbers of components varied from 1 to 18, showing the appearance of new biochemical components~(red numbers in circles) and their splitting/mixing. The degree of component similarity between two analysis with subsequent numbers of components is indicated by solid (strong), dashed (medium), and dotted (weak) lines. 
} \label{Fig3}
\end{figure*}

\subsection{Noise filtering: SVD-ADC}\label{Sec:Results1}
As well-documented in the literature, conventional SVD procedure for spectral image processing allows to separate signal from noise by rejecting small singular values and their corresponding singular vectors from the further analysis. SVD-ADC approach, proposed in this paper, utilises autocorrelation coefficients of spatial and spectral singular vectors for noise filtering. In particular, for each pair of singular vectors we calculate spatial and spectral autocorrelation coefficients and discard the singular vectors, which have a mean autocorrelation coefficient smaller than 50\%. 

\Fig{Fig2}{D} shows the autocorrelation coefficients map for singular vectors received from SVD-ADC analysis of 6 Raman images. The dashed diagonal line represents a decision line for mean autocorrelation coefficients at $R_\mathrm{thr}=50\%$ cut-off, that separates the coefficients above the line~(meaningful components) from those below it~(noise). Note that read-noise and shot noise are localized around zero, whereas physically meaningful components show high mean autocorrelation.

Importantly, the new approximating matrix, received after SVD on given data, is compressed~(or has \textit{significantly} reduced-rank), therefore it is important to compromise the cut-off value and avoid rejection of meaningful components. For our data example, the SVD-ADC with a 50\,\% cut-off of mean autocorrelation coefficients has found the approximating matrix, consisting of 19~singular components. To verify this number, we show the spatial distributions and corresponding spectra for two singular vectors with mean autocorrelation coefficients, which are 20\% lower and higher than 50\,\% value~(\Fig{Fig2}E,F). 
For a pair of \textit{discarded} singular vectors, corresponding to 30\,\% mean autocorrelation, spatial maps do not show any recognised pattern and spectrum resembles noise, whereas for a pair of meaningful singular vectors with 70\,\% mean autocorrelation, spatial maps show some co-localized features and spectrum exhibits several clear Raman bands.

\subsection{Background subtraction: BGF}\label{Sec:Results2}

\Fig{Fig2}B,C shows an example of our BGF background subtraction algorithm applied to the Raman data, found from the SVD-ADC as detailed in the previous section. The
Raman spectra at two spatial pixels (high and low background) from the selected hyperspectral image \textit{before}~(black line) and \textit{after}~(green line) background removal with the representative bottom Gaussian fit~(red line) are shown.

\subsection{Q-US/PS-NMF unmixing procedure}\label{Sec:Results3}
After SVD-ADC filtering and BGF background subtraction, we analysed the Raman data using the Q-US/PS-NMF in the fingerprint (370\,--\,1783\,cm$^{-1}$) region. 
The Q-US/PS-NMF analysis found that the given 6 Raman images were optimally described using 11 separate components~($\Comp$). To validate this component number selection, we performed the Q-US/PS-NMF with the numbers of components varied from 1 to 18, and then investigated the evolution of resulting component spectra (see \Fig{Fig3}D), which includes the formation of new biochemical components~(red numbers in circles) and their splitting/mixing. The degree of component similarity between two analyses with subsequent numbers of components is indicated by solid (strong), dashed (medium), and dotted (weak) lines. For example, the formation of \textbf{$\Comp_{1}$} and \textbf{$\Comp_{10}$} in 11-component analysis, which will be later attributed to actin and hydroxyapatite, respectively, was a result of splitting of \textbf{$\Comp_{9}$} in 10-component analysis.
Note the components are sorted by their mean concentration, from largest to smallest.
The dependence of the factorization error on the number of components was also inspected. The results (see Fig.\,S5 of the Supplementary Material) show a gradual decrease of this error when increasing the number of components, and therefore imply that the factorization error can not be used to determine the optimal number of components in the Q-US/PS-NMF data analysis.

7 components were found to be spectrally assigned to actin~(\textbf{$\Comp_{1}$}, HQI 85\,\%); elastin~(\textbf{$\Comp_{3}$}, HQI 84\,\%); a mixture of phospholipids and triglycerides~(\textbf{$\Comp_{4}$}, HQI 93\,\%) at a ratio of about 2:1 (\Fig{Fig3}A); a mixture of cholesterol, cholesteryl esters with saturated fatty acid chains, and oxidized all-\textit{trans} retinol also known as vitamin A~(\textbf{$\Comp_5$}, HQI 94\,\%) at ratios of about 6:5:1 (\Fig{Fig3}B); collagen~(\textbf{$\Comp_{6}$}, HQI 83\,\%); hydroxyapatite~(\textbf{$\Comp_{10}$}, HQI 93\,\%); a mixture of $\beta$-carotene and cholesterol~(\textbf{$\Comp_{11}$}, HQI 88\,\%) at a ratio of about 3:1~(see the component spectra in Figs. S6-S16 of the Supplementary Material). 
Here, the contributions of biochemical reference components into each component spectra was determined using NLS fitting algorithm. The degree of similarity between each component spectra and its biochemical model was indicated by a hit quality index~(HQI). The model uses the Raman spectra of the chemical species of analytical standards\,\citep{You2017, Czamara2015, Krafft2005}. Additionally, $\Comp_{2}$ was found to be consistent with the Raman spectrum of water~(HQI 97\,\%) using an ID expert tool of Bio-Rad's KnowItAll Vibrational Spectroscopy software with Raman Spectral Libraries.

Generally, the results of the Q-US/PS-NMF analysis are consistent with the published results found from the VCA analysis of the same samples\,\citep{You2017}. However, our  method allowed to retrieve additional biochemical information on the aortic tissue composition. In particular, it reveals a new biochemical component, which is spectrally attributed to oxidized linoleic acid~(\textbf{$\Comp_{8}$}, HQI 90\,\%) (\Fig{Fig3}C). Importantly, the spectrum of this component (blue line) shows two unique Raman bands at 870~cm$^{-1}$ and 1744~cm$^{-1}$, assigned to the -O-OH stretching mode, characteristic of hydrogen peroxide (magenta line in inset), and C=O stretching vibrations, which both occur in the process of lipid peroxidation. Altogether, these observations support that linoleic acid is oxidatively modified in the atherosclerotic aortic tissue.

Furthermore, the Q-US/PS-NMF enabled to identify the signatures of oxidized all-\textit{trans} retinol (cyan line) in the component spectra~\textbf{$\Comp_5$} (blue line), composed mainly of cholesterol (green line) and cholesteryl esters (magenta line) as shown in the inset of \Fig{Fig3}B. The appearance of a Raman band at 1529~cm$^{-1}$ as well as a broad shoulder with peaks at 1618~cm$^{-1}$ and 1635~cm$^{-1}$ reflect oxidation-induced structural changes of all-\textit{trans} retinol, ensuing in the process of its degradation\,\citep{Failloux2003}, and therefore suggests its attribution to oxidized all-\textit{trans} retinol. Importantly, the Raman feature at 1635~cm$^{-1}$ in the \textbf{$\Comp_5$} can be also assigned to the C=O stretch of ketones, previously observed in the spectrum of oxidized low density lipoprotein\,\citep{Wang2003}. One more example of new information deduced from the Q-US/PS-NMF is the presence of cholesterol in the $\beta$-carotene component spectrum~(\textbf{$\Comp_{11}$}), which indicates that these species form a tightly-connected structure in the aortic tissue.

Notably, we also observed that \textbf{$\Comp_{9}$} might be represented as a mixture of triglycerides and arachidonic acid at a ratio of about 3:1, but with a low HQI value of 61\,\%, which might indicate that these components are strongly modified in the atherosclerotic tissues due to oxidation.

To claim we can quantify the information from the Raman images using the Q-US/PS-NMF, we show in \Fig{Fig3}A-C the spatially-resolved maps of the concentrations of $\Comp_{4}$, $\Comp_{5}$, and $\Comp_{8}$ over the investigating images, representing atherosclerotic aortic tissue regions~(A1, A2, A3) and non-atherosclerotic controls~(C1, C2, C3). The similar figures of the remaining components are shown in Supplementary Material. 
The concentration maps of these components show that oxidized linoleic acid~($\Comp_{8}$) aggregates are strongly co-localized with cholesterol/saturated cholesteryl ester crystals~($\Comp_{5}$) in the atherosclerotic plaques as well as  with oxidized lipid accumulations~($\Comp_{4}$) enriched with triglyceride and phosphocholine in the foam cells. This oxidized linoleic acid component shows by one order of magnitude increased concentrations in the atherosclerotic plaque lesions compared  to the atherosclerotic plaque-negative and non-atherosclerotic control regions. Altogether, this observation indicates the interplay between oxidized linoleic acid and cholesterol and triglyceride- rich components in the aortic tissue, which might be one of the possible molecular mechanisms involved in the evolution of the atherosclerotic plaque. The studies, which used several different animal models of atherosclerosis, also found that oxidized fatty acids in the diet increase fatty streak lesion formation in the aortic tissues\,\citep{Staprans2005}.

Altogether, we have demonstrated that the new developed Q-HIU method presented in this paper is capable to quantitatively decompose the spatially-resolved Raman spectra of human atherosclerotic aortic tissues into a number of individual biochemical components such as structural proteins (actin, elastin, collagen), oxidatively modified lipids, and minerals, which are characteristic bio-markers of atherosclerosis.

\section{Conclusion}
To conclude, we present a new efficient Quantitative Hyperspectral Image Unmixing (Q-HIU) method, allowing to autonomously analyse large-scale Raman micro-spectroscopy data with minimum input parameters and high accuracy. This in-house developed method implemented in MATLAB integrates three consecutive steps of data analysis, called Singular Value Decomposition with innovative Automatic Divisive Correlation, Bottom Gaussian Fitting, and an efficient Quantitative Unsupervised/Partially~Supervised Non-negative Matrix Factorization method, which is capable to operate as a partially supervised method, when one or several samples' constituents are known. This is implemented by fixing the spectra of the known compounds in the approximating factorization expansion of the data matrix. 
The great advantage of the Q-US/PS-NMF algorithm to directly work with matrices allows to achieve order-of-magnitude acceleration in computational speed as compared with the popular realization of MCR-ANLS. The Q-HIU method was validated on large volume Raman images of atherosclerotic tissues, showing a proof-of-principle for the biomolecular characterisation and quantitative imaging of individual biochemical components(e.g.: cholesterol/cholesteryl ester, oxidized linoleic acid, calcium hydroxyapatite, $\beta$-carotene crystals) of atherosclerotic plaques compared to non-atherosclerotic controls. 

\section*{Acknowledgements}
The authors acknowledge S.G.~Tikhodeev, N.A.~Gippius, P.~Borri, W.~Langbein, K.~Triantafilou, and F.~Masia for useful discussions.

\section*{Funding}
This work has been supported by the Cardiff University College of Biomedical and Life Sciences under the International Scholarship for PhD student and RFBR Project No. 16-29-03283.

\section*{Data Availability}
The software used to produce the results of this paper and containing three functions: SVD-ADC, BGF, and Q-US/PS-NMF, is freely available on the web at \url{https://github.com/LobanovaEG-LobanovSV/Q-HIU.git}.

\pagebreak
\widetext
\begin{center}
\textbf{\large Supplementary Material: Efficient quantitative hyperspectral image unmixing method for large-scale Raman micro-spectroscopy data analysis}
\end{center}
\setcounter{equation}{0}
\setcounter{figure}{0}
\setcounter{table}{0}
\makeatletter
\renewcommand{\theequation}{S\arabic{equation}}
\renewcommand{\thefigure}{S\arabic{figure}}

\section{Euclidean scalar product}
Euclidean scalar product of two arbitrary vectors $V_i(\xi)$ and $V_j(\xi)$, where $\xi$ consists of a finite number of discrete points $\{\xi_k\}$, is 
\begin{equation}
V_i(\xi)\cdot V_j(\xi) = \sum_k V_i(\xi_k) V_j(\xi_k),\label{Eq:ESP_1}
\end{equation}
where summation is performed over all elements of the set $\{\xi_k\}$.
This equation can be also rewritten in a matrix form
\begin{equation}
V_i(\xi)\cdot V_j(\xi) = \sum_k V_{ik}V_{jk}.\label{Eq:ESP_2}
\end{equation}

In \Eqs{Eq:ESP_1}{Eq:ESP_2}, $V_i(\xi)$ denotes $C_i(\r)$ or $S_i(\nu)$, i.e.
\begin{gather}
C_i(\r)\cdot C_j(\r) = \sum_{k=1}^{N_p} C_i(\r_k) C_j(\r_k) = \sum_{k=1}^{N_p} C_{ik}C_{jk},\\
S_i(\nu)\cdot S_j(\nu) = \sum_{k=1}^{N_s} S_i(\nu_k) S_j(\nu_k) = \sum_{k=1}^{N_s} S_{ik}S_{jk},
\end{gather}

\section{Correlation coefficient}
Correlation coefficient of two arbitrary vectors $\A$ and $\B$ is 
\begin{equation}
R\left[\A,\B\right] = \frac{|\sum_k(A_k-\overline{A})(B_k-\overline{B})|}{\sqrt{\sum_k(A_k-\overline{A})^2}\sqrt{\sum_k(B_k-\overline{B})^2}},
\end{equation}
where $\overline{A}$ and $\overline{B}$ represent mean values of $\A$ and $\B$, respectively, i.e.
\begin{gather}
\overline{A}=\frac{1}{N}\sum_{k=1}^N A_k,\\
\overline{B}=\frac{1}{N}\sum_{k=1}^N B_k.
\end{gather}
Here, $N$ indicates a number of elements in the vectors $\A$ and $\B$.

\section{Comparison of Q-US/PS-NMF with other software}
\begin{figure}
\begin{center}
\includegraphics[width=0.5\linewidth]{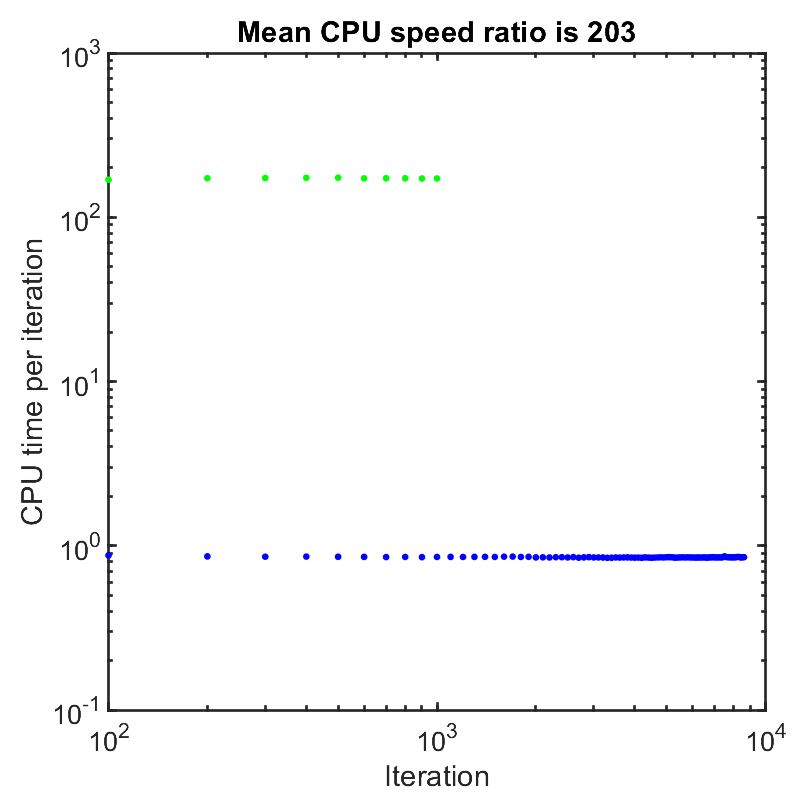}
\end{center}
\caption{
Comparison of computational efficiency of the MCR-ALS by \citep{Jaumot2015} (green dots) with Q-US/PS-NMF (blue dots).
}  
\label{FigS1}
\end{figure}
We compared computational efficiencies of Q-US/PS-NMF, MCR-ALS by \citep{Jaumot2015}, and FC-NMF by \citep{Kim2007a} on a computer with an Intel Core i7-4770K processor, 32 GB of RAM, and Ubuntu operating system.
For the comparison we used real experimental large-scale Raman micro-spectroscopy images ($\simeq${}$5\cdot10^{5}$-by-$10^3$ matrix $\I$) from human thoracic aortic tissue in atherosclerosis [samples A1~(022$\_$p$\_$000), A2~(025$\_$p$\_$000), A3~(031$\_$p$\_$005)] together with non-atherosclerotic age-matched controls [samples C1~(043$\_$000), C2~(046$\_$000), C3~(047$\_$002)], which were sourced from\,\citep{You2017,You2017b} and preprocessed using newly developed SVD-ADC and bottom Gaussian fitting subtraction.
$N=11$ components were used in factorization.

MCR-ALS by \citep{Jaumot2015} was downloaded on 01.04.2018 from \url{https://mcrals.wordpress.com/download/mcr-als-2-0-toolbox}.
The functions "alsOptimization.m" and "fnnls.m" were used for the factorization.
\Fig{FigS1}{} shows comparison of MCR-ALS by \citep{Jaumot2015} (green dots) with Q-US/PS-NMF (blue dots).
One can see from this figure that Q-US/PS-NMF is 200 times faster than MCR-ALS by \citep{Jaumot2015}.

\begin{figure}
\begin{center}
\includegraphics[width=0.5\linewidth]{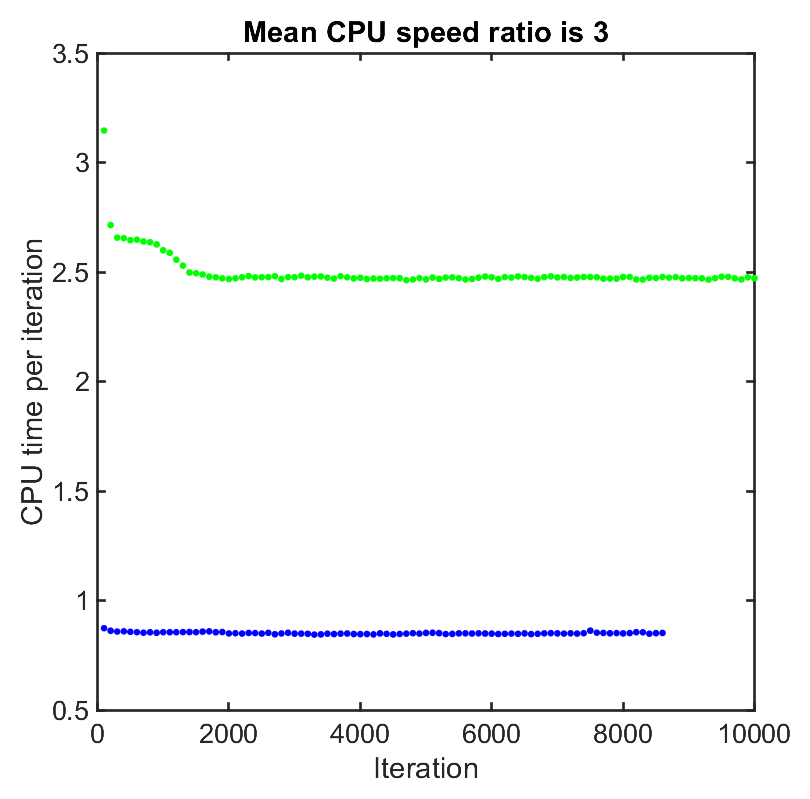}
\end{center}
\caption{
Comparison of computational efficiency of the FC-NMF algorithm by Kim and Park (green dots) with the Q-US/PS-NMF developed by us (blue dots).
}  
\label{FigS2}
\end{figure}

FC-NMF implementation by \citep{Kim2007a} was downloaded on 01.04.2018 from  \url{https://github.com/kimjingu/nonnegfac-matlab}.
The function "nmf.m" with default parameters (except tolerance and number of iterations) was used for the factorization:
\begin{verbatim}
[W,H,iter,REC] = ...
     nmf(I,11,'tol',1e-16,'max_iter',20000);
\end{verbatim}
The variables "REC.HIS.iter" and "REC.HIS.elapsed" were used to produce \Fig{FigS2}{} showing comparison of FC-NMF by \citep{Kim2007a} (green dots) with Q-US/PS-NMF (blue dots).
One can see from this figure that Q-US/PS-NMF is 3 times faster than FC-NMF by \citep{Kim2007a}.

\section{Convergence of Q-US/PS-NMF to local/global minima for the data presented in the article}

\begin{figure}
\begin{center}
\includegraphics[width=0.5\linewidth]{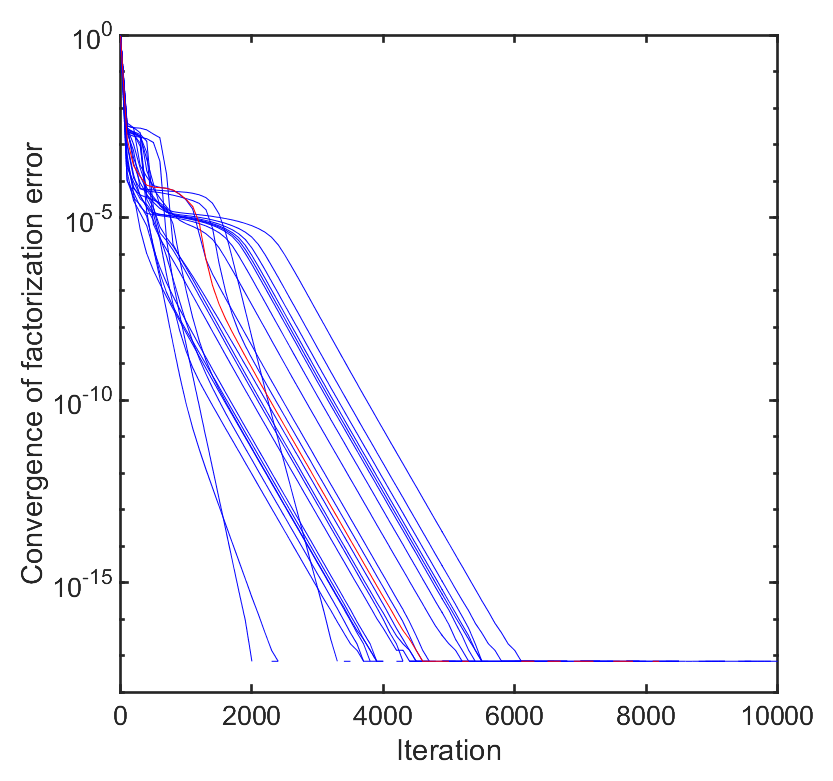}
\end{center}
\caption{
Dependence of relative factorization error on iteration.
The red line indicates the replication with minimum final factorization error.
}  
\label{FigS3}
\end{figure}

\begin{figure}
\begin{center}
\includegraphics[width=0.5\linewidth]{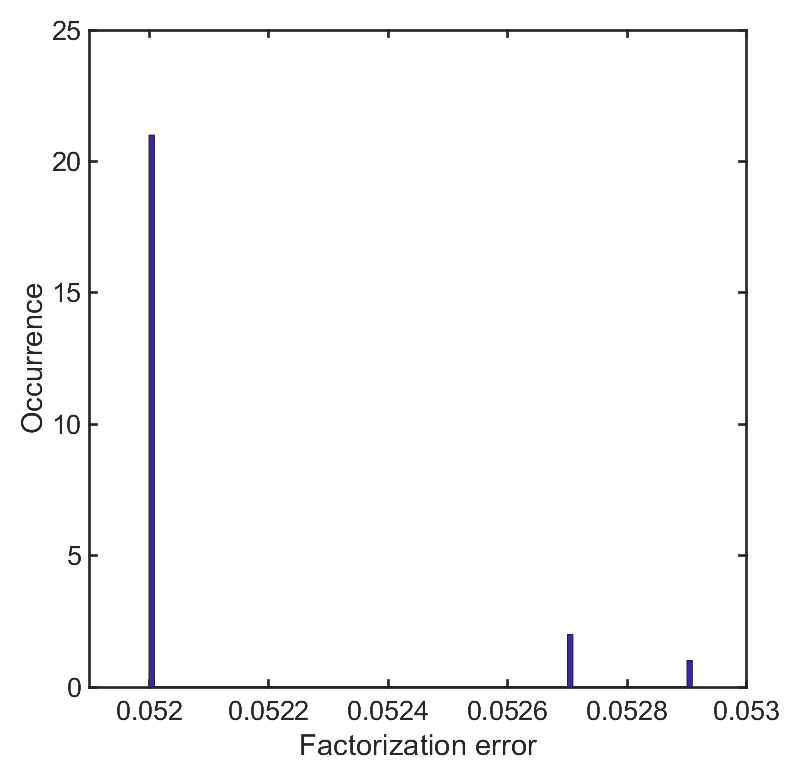}
\end{center}
\caption{
Occurrence of local minima after $N_{repl}=24$ replications of the ANLS procedure.
}  
\label{FigS4}
\end{figure}

In order to find a local minimum of a relative factorization error
\begin{equation}
E(\C,\S)=\frac{\|\I-\C^\mathrm{T}\S\|_\mathrm{F}}{\|\I\|_\mathrm{F}},
\end{equation}
we performed $N_{iter}=20000$ iterations.
In order to find a global minimum, we repeat ANLS procedure $N_{repl}=24$ times with random initial matrix $\S_0$.
The results of calculation for the number of components $N=11$ is shown in \Fig{FigS3}.
In this figure, factorization error after each iteration $E_i=E(\C_i,\S_i)$ relative to the final factorization error $E_{20000}$ is shown by blue lines, i.e. $\Delta E_i = E_i-E_{20000}$ is shown.
The red line indicate the replication with minimum final factorization error.
One can see from the figure that Q-US/PS-NMF converges exponentially to a local minimum and $N_{iter}=20000$ iterations is sufficient.

Occurrence of local minima is show in \Fig{FigS4}.
One can see from the figure that even several replications is sufficient to find the global minimum.

\section{Results of the Q-HIU analysis}

\begin{figure}
\begin{center}
\includegraphics[width=0.5\linewidth]{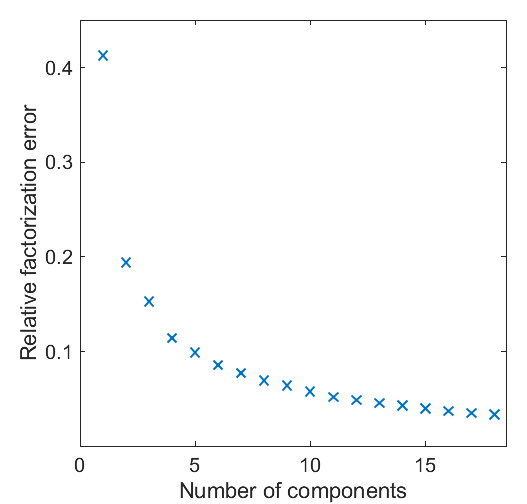}
\end{center}
\caption{
Dependence of the relative factorization error $E$ on the number of components $N$.
}  
\label{FigS5}
\end{figure}

Dependence of the relative factorization error $E$ (i.e. the final relative factorization error $E_{20000}$) on the number of components $N$ is shown in \Fig{FigS5}.

\begin{figure*}[tb]
\includegraphics[width=0.38\linewidth]{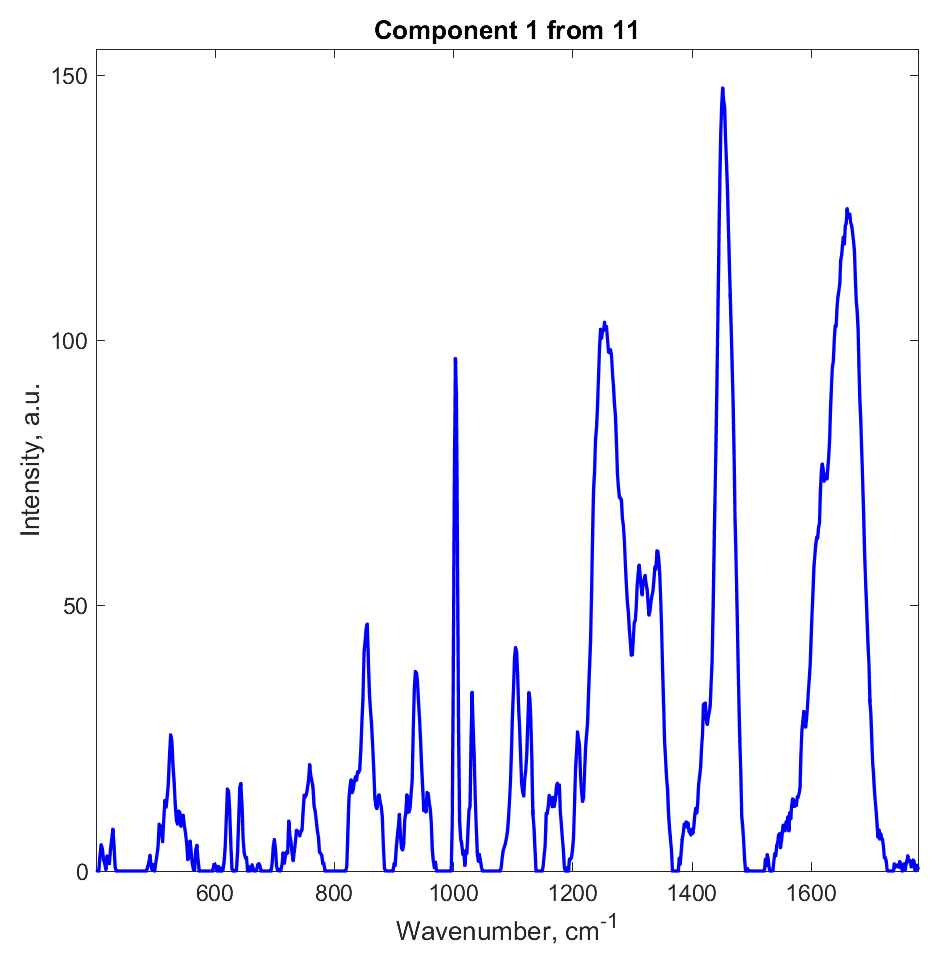}
\includegraphics[width=0.50\linewidth]{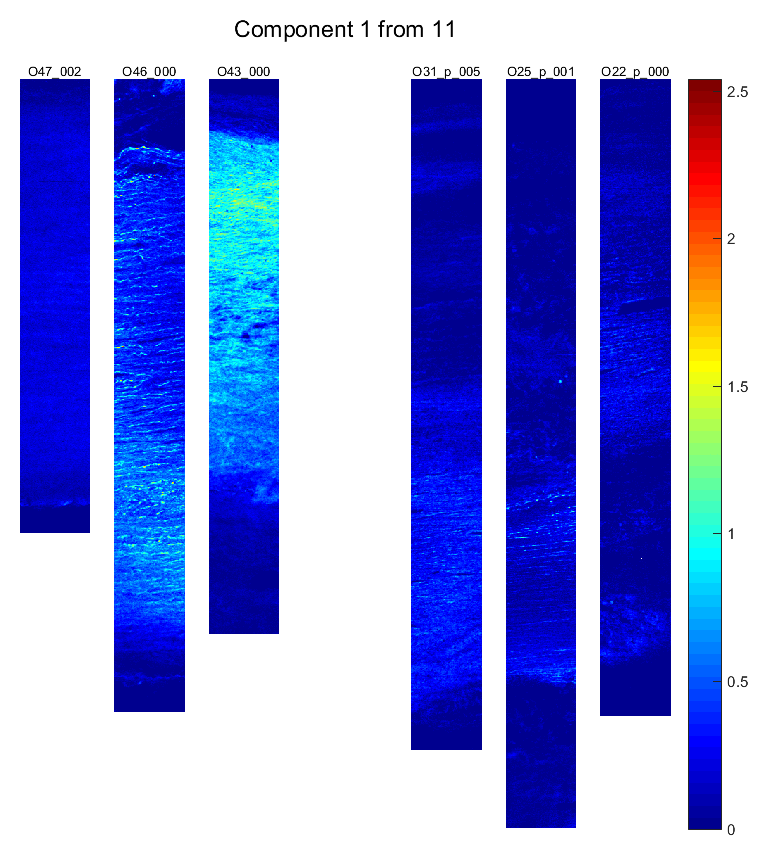}
\caption{Component spectra~(left) and spatial distributions of concentration~(right) of $\Comp_{1}$ found from the Q-US/PS-NMF of 6 Raman images from atherosclerotic aortic tissue regions~(A1, A2, A3) and non-atherosclerotic controls~(C1, C2, C3).}\label{Fig:Comp1}
\end{figure*}

Spectra and concentration maps of the chemical components, retrieved from 6 Raman images of atherosclerotic aortic tissue regions~[samples A1~(022$\_$p$\_$000), A2~(025$\_$p$\_$000), A3~(031$\_$p$\_$005)] and non-atherosclerotic controls~[samples C1~(043$\_$000), C2~(046$\_$000), C3~(047$\_$002)] using Q-HIU with $\sigma=300$ cm$^{-1}$ and $N=11$, are presented in \Figs{Fig:Comp1}{Fig:Comp11}.

\begin{figure*}[tb]
\includegraphics[width=0.38\linewidth]{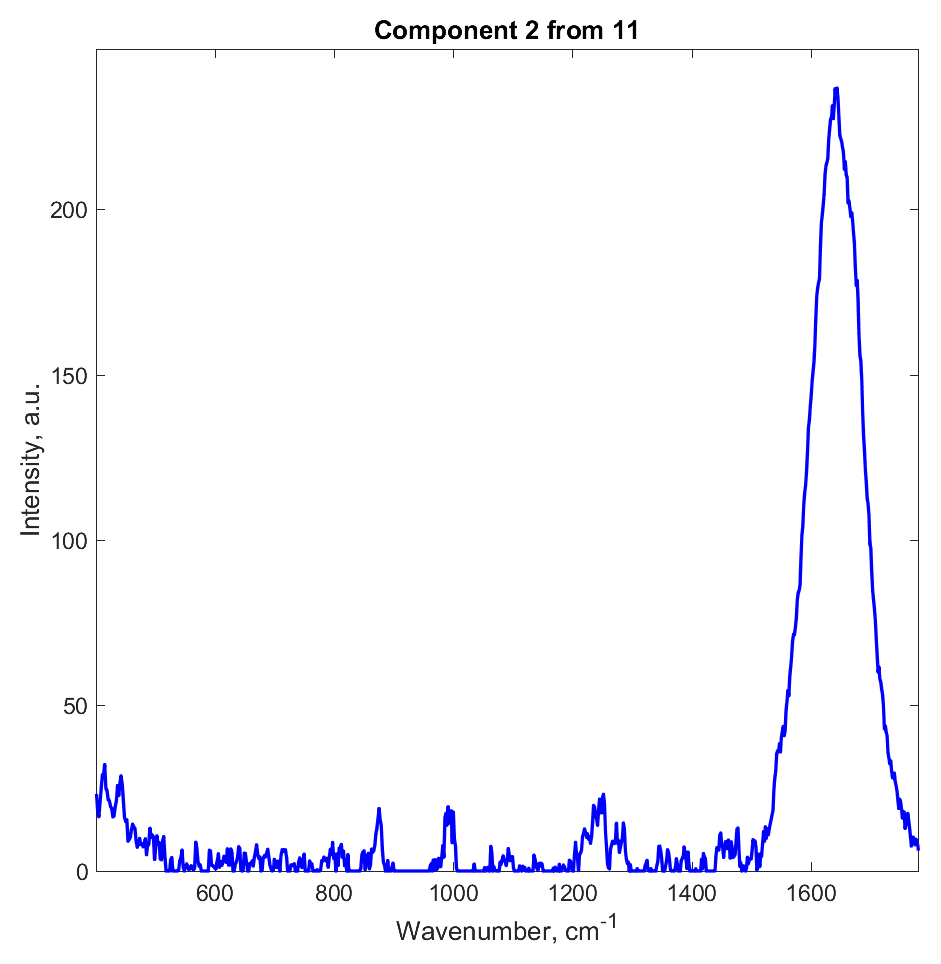}
\includegraphics[width=0.50\linewidth]{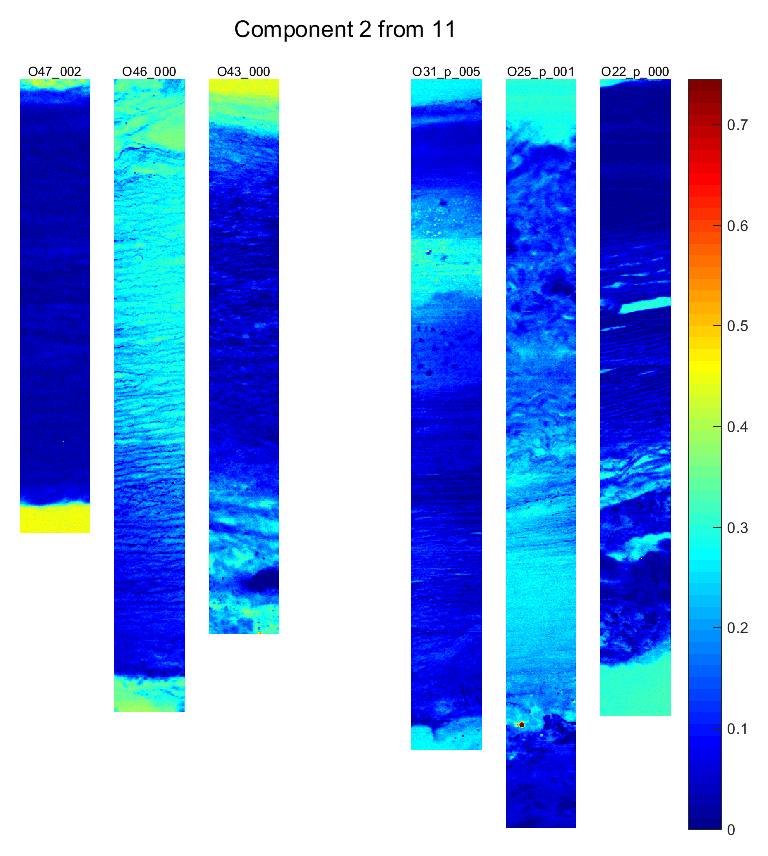}
\Capttwo{2}
\end{figure*}

\begin{figure*}[tb]
\includegraphics[width=0.38\linewidth]{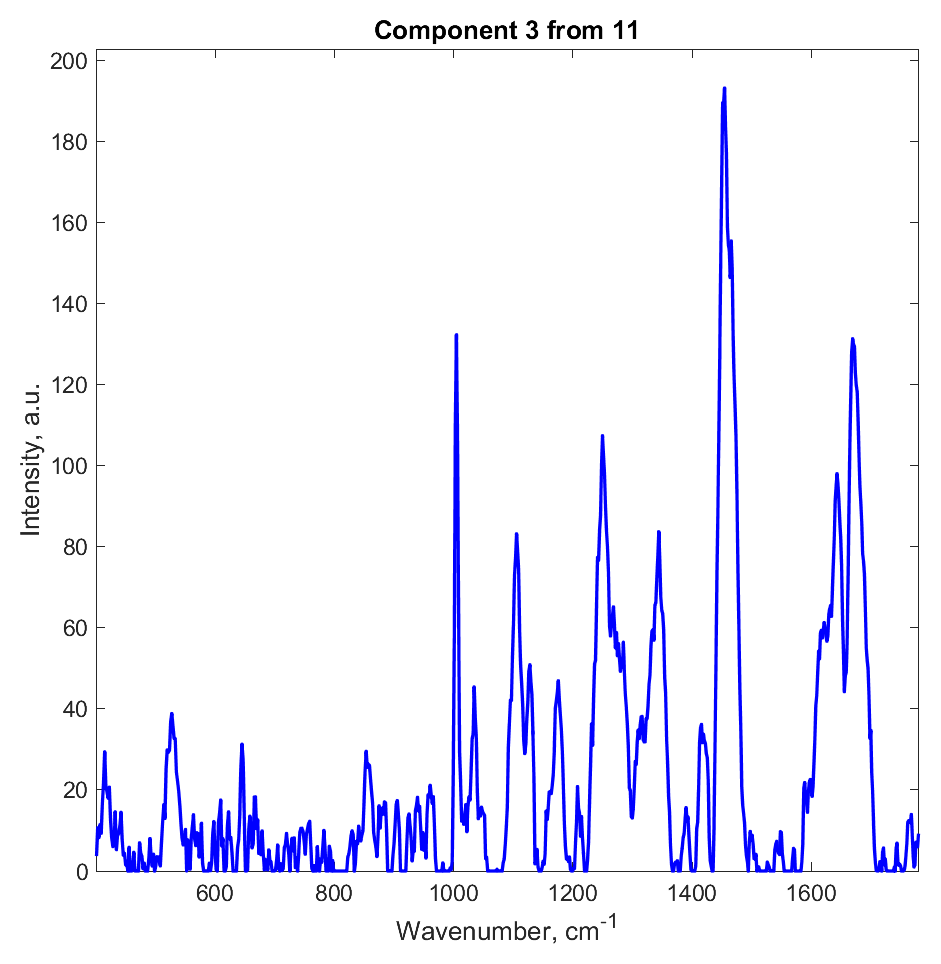}
\includegraphics[width=0.50\linewidth]{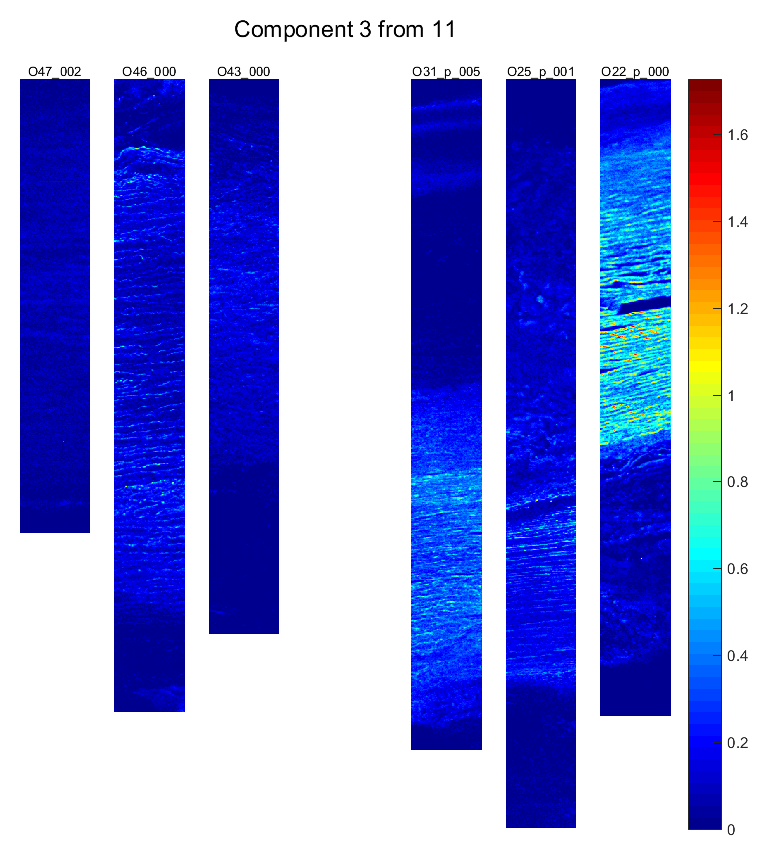}
\Capttwo{3}
\end{figure*}

\begin{figure*}[tb]
\includegraphics[width=0.38\linewidth]{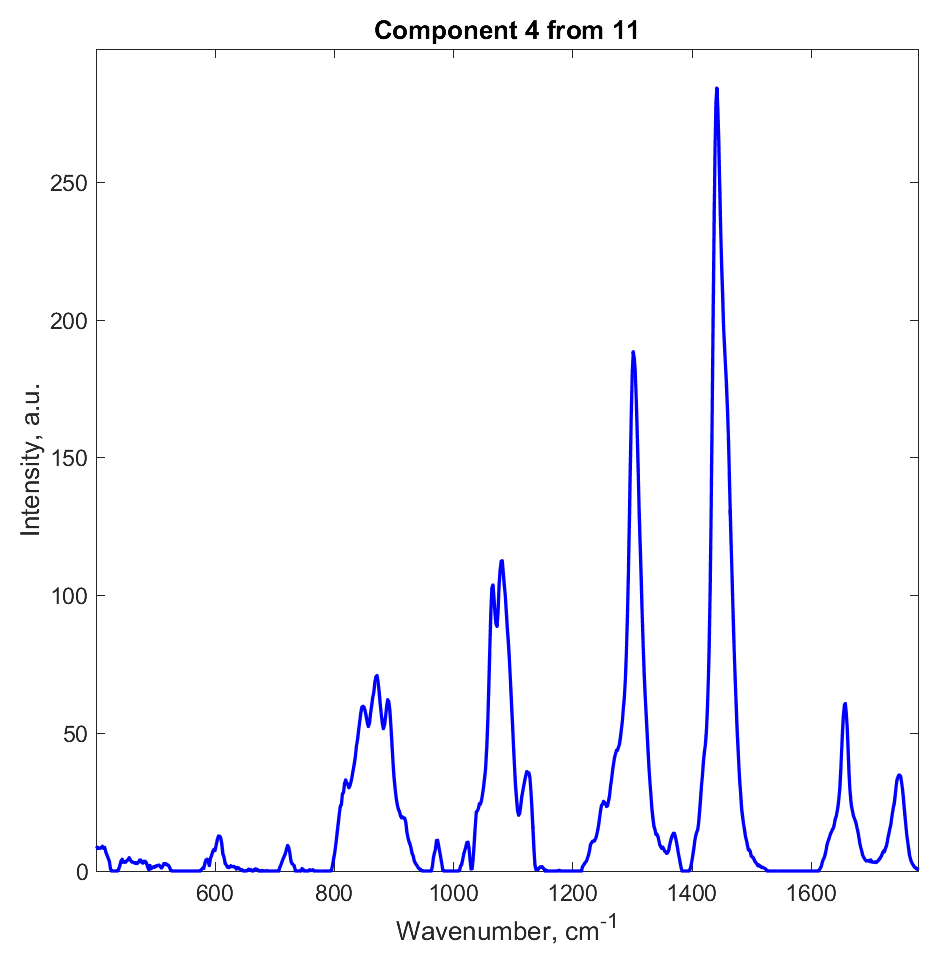}
\includegraphics[width=0.50\linewidth]{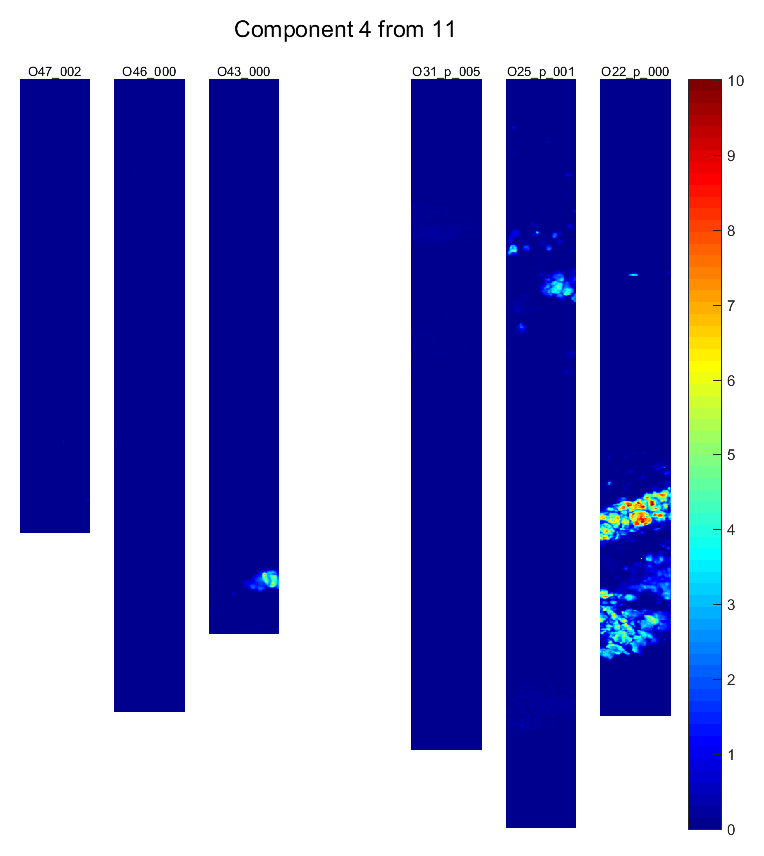}
\Capttwo{4}
\end{figure*}

\begin{figure*}[tb]
\includegraphics[width=0.38\linewidth]{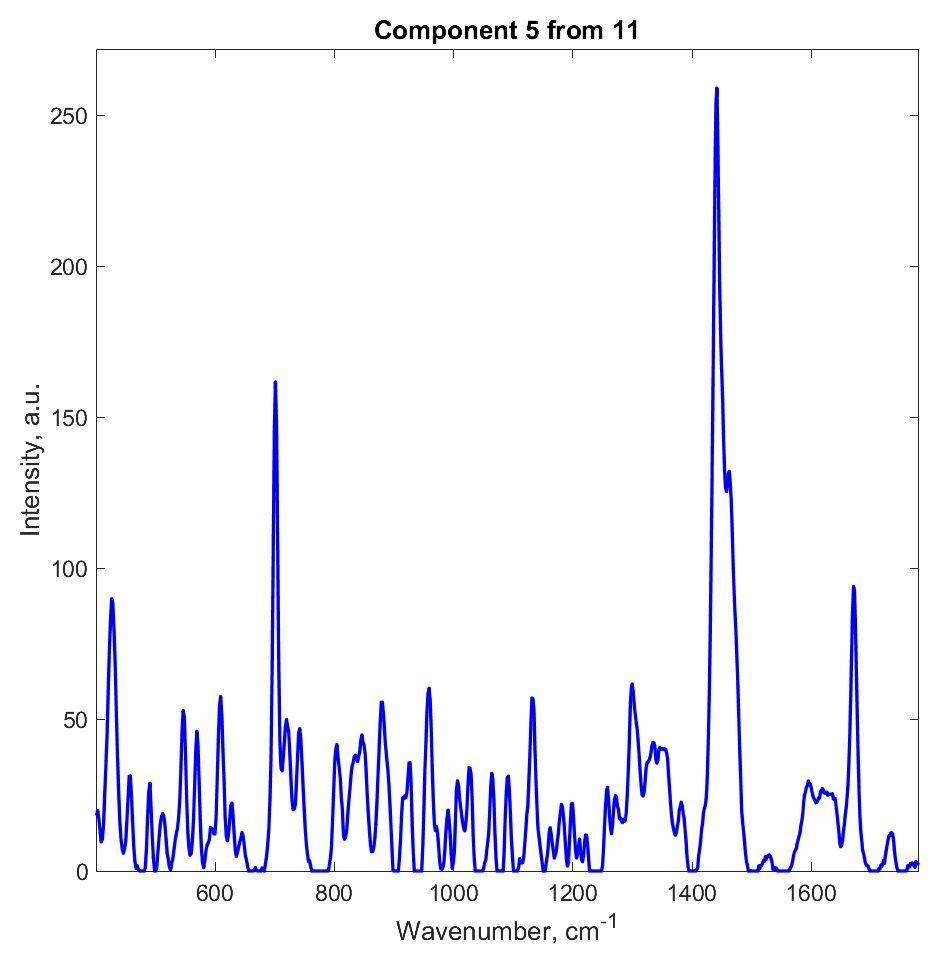}
\includegraphics[width=0.50\linewidth]{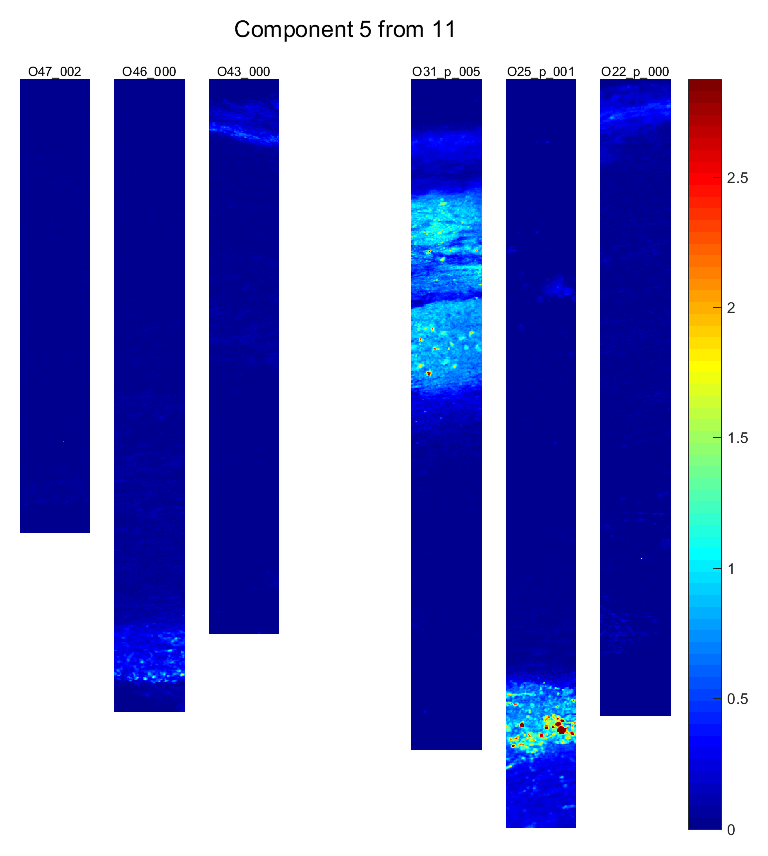}
\Capttwo{5}
\end{figure*}

\begin{figure*}[tb]
\includegraphics[width=0.38\linewidth]{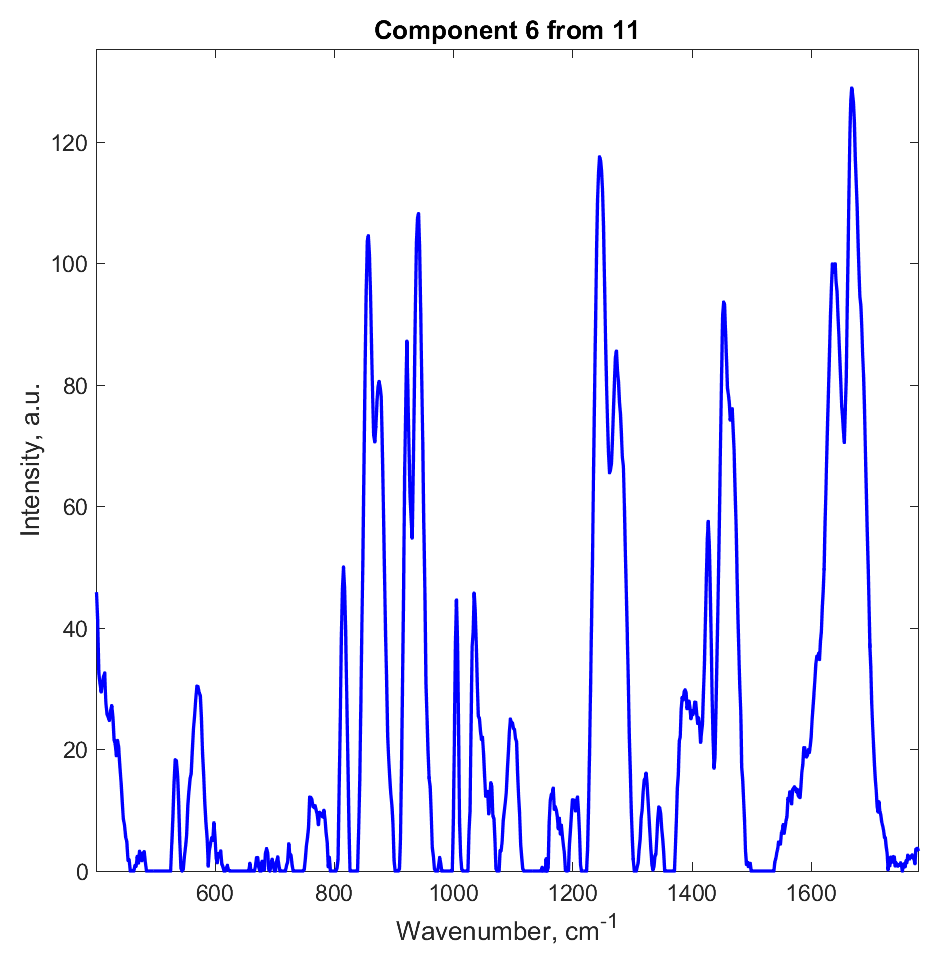}
\includegraphics[width=0.50\linewidth]{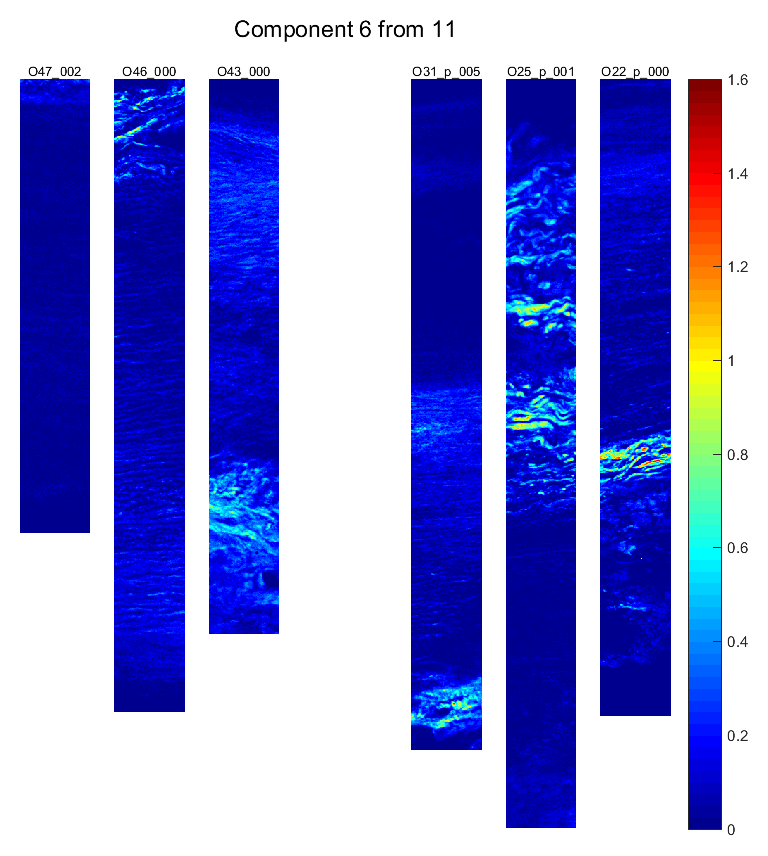}
\label{Tc_vs_N}
\Capttwo{6}
\end{figure*}

\begin{figure*}[tb]
\includegraphics[width=0.38\linewidth]{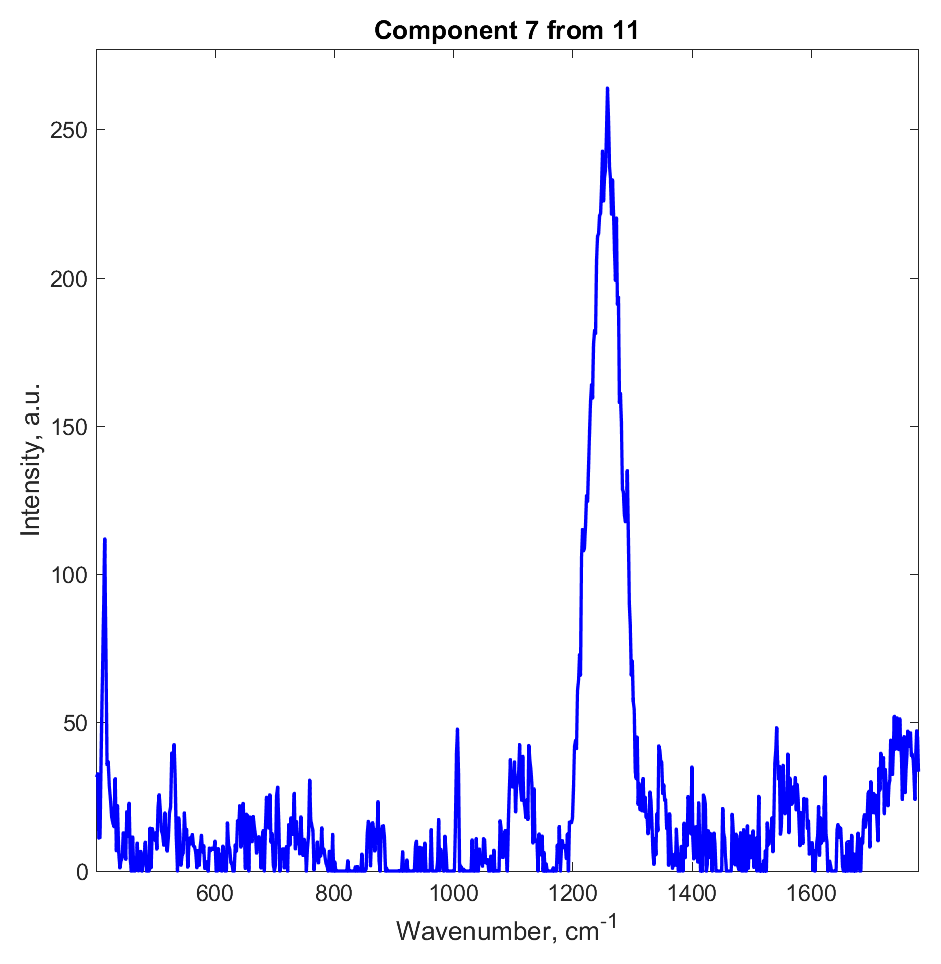}
\includegraphics[width=0.50\linewidth]{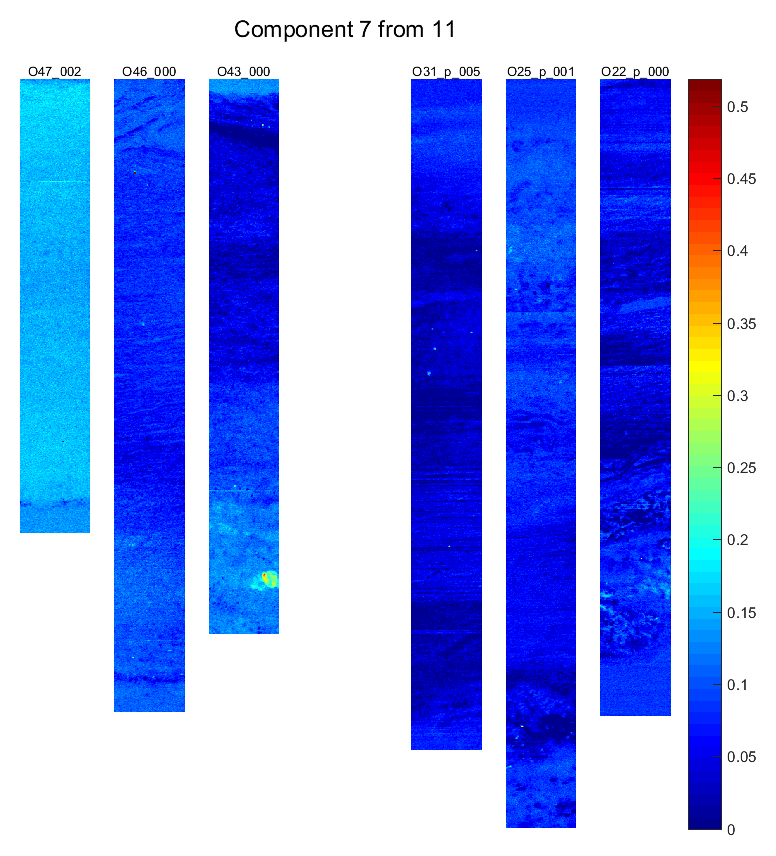}
\Capttwo{7}
\end{figure*}

\begin{figure*}[tb]
\includegraphics[width=0.38\linewidth]{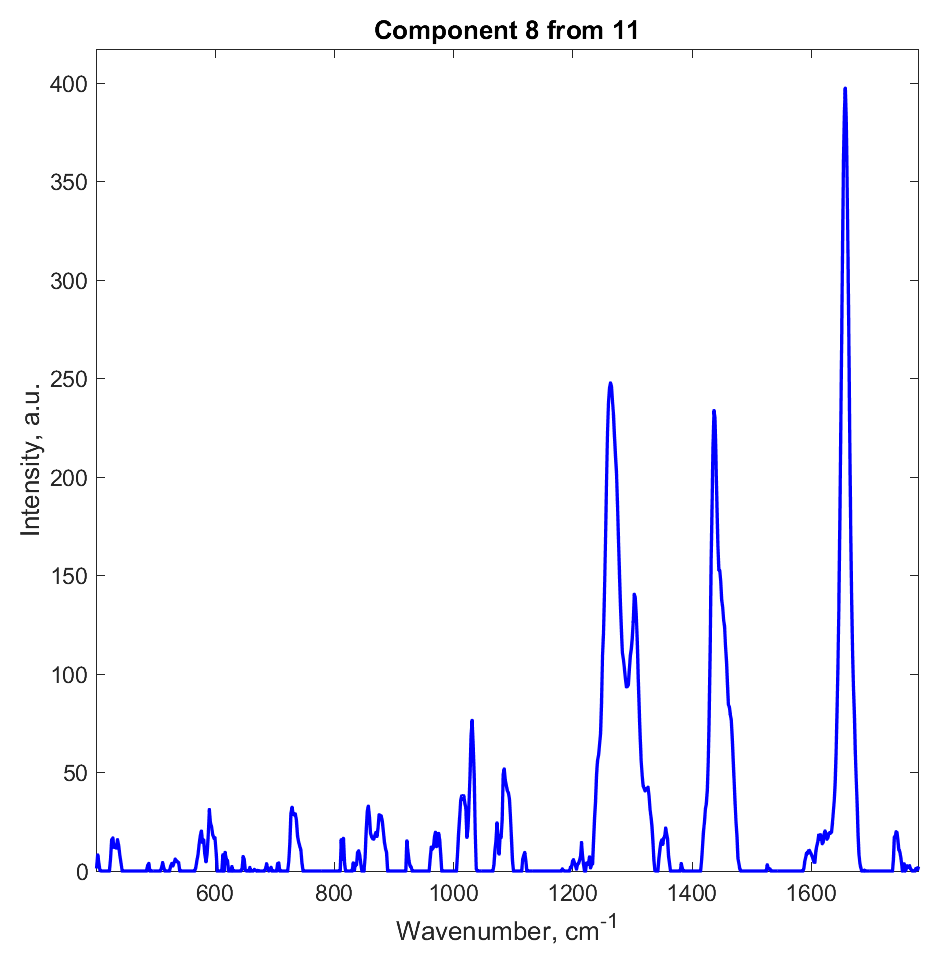}
\includegraphics[width=0.50\linewidth]{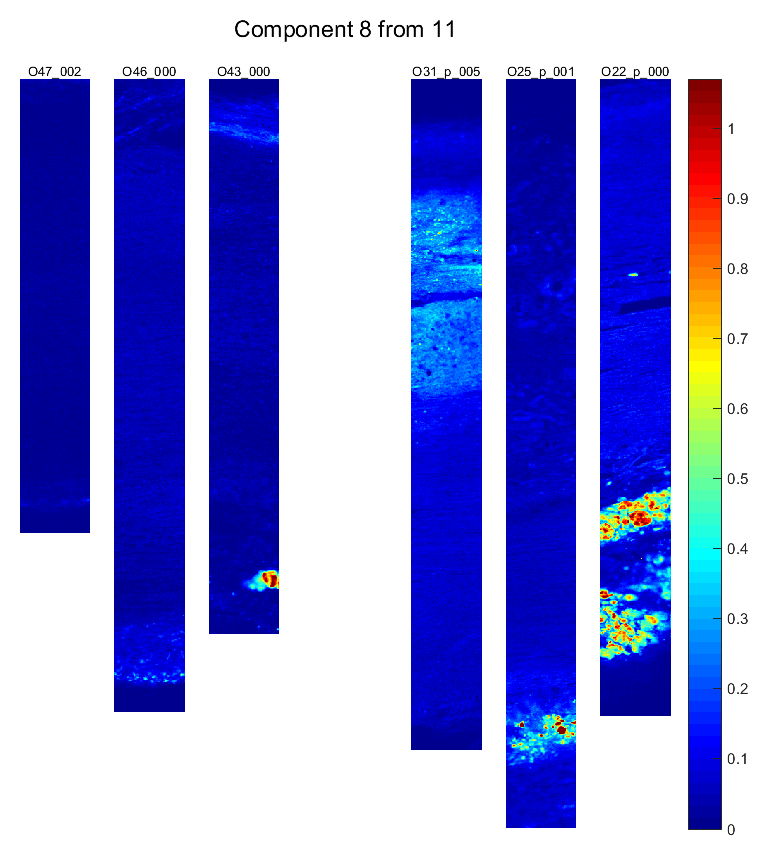}
\Capttwo{8}
\end{figure*}

\begin{figure*}[tb]
\includegraphics[width=0.38\linewidth]{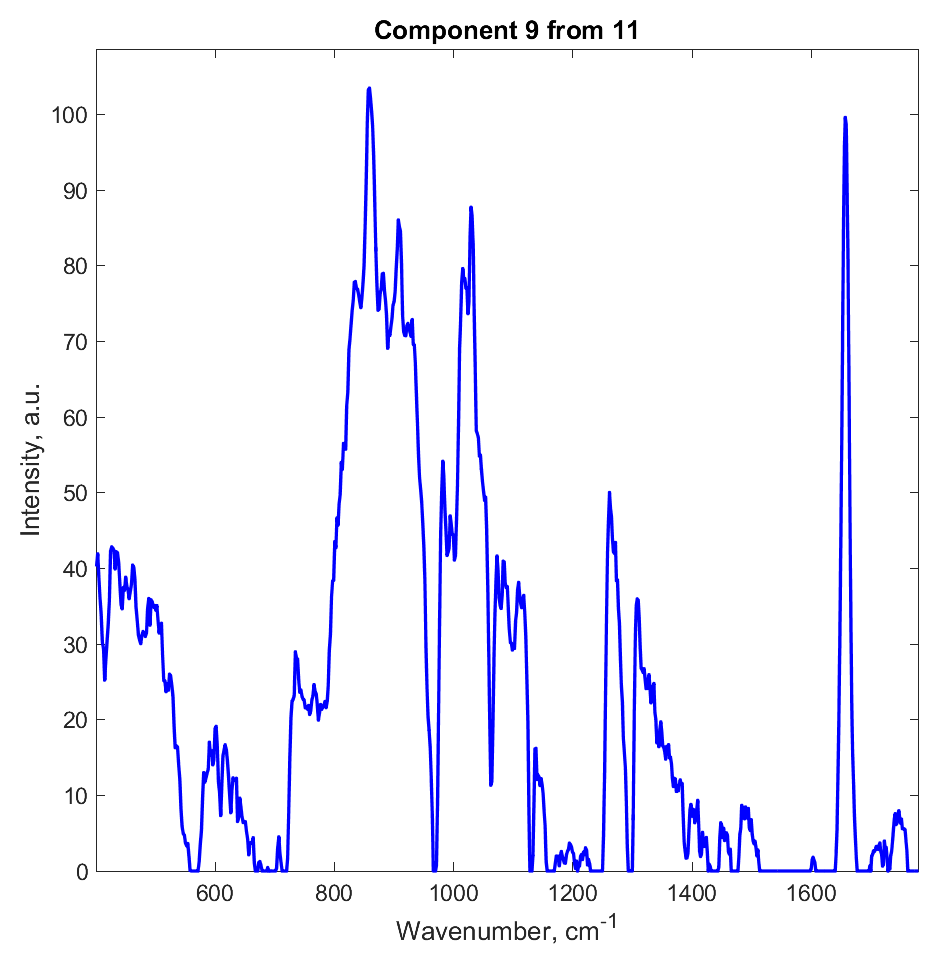}
\includegraphics[width=0.50\linewidth]{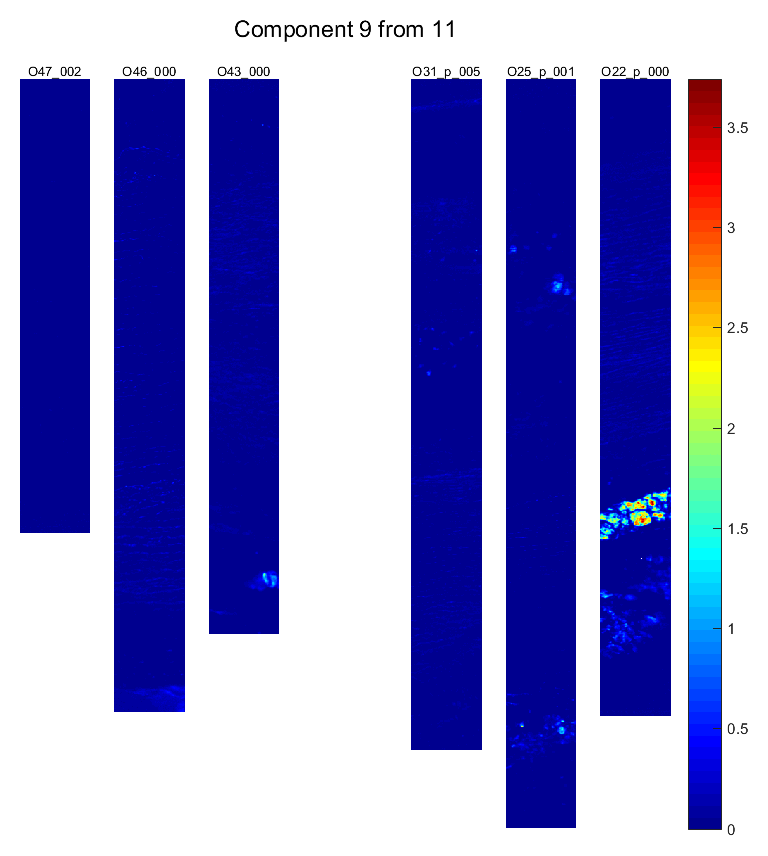}
\Capttwo{9}
\end{figure*}

\begin{figure*}[tb]
\includegraphics[width=0.38\linewidth]{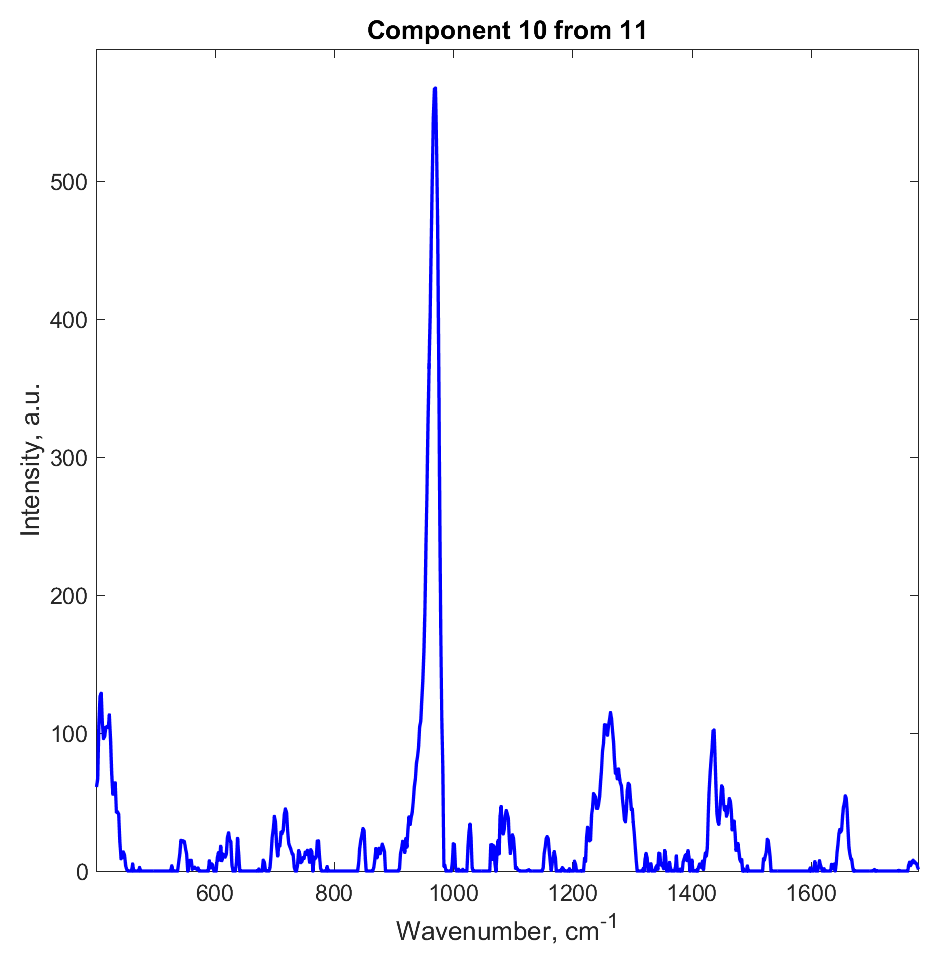}
\includegraphics[width=0.50\linewidth]{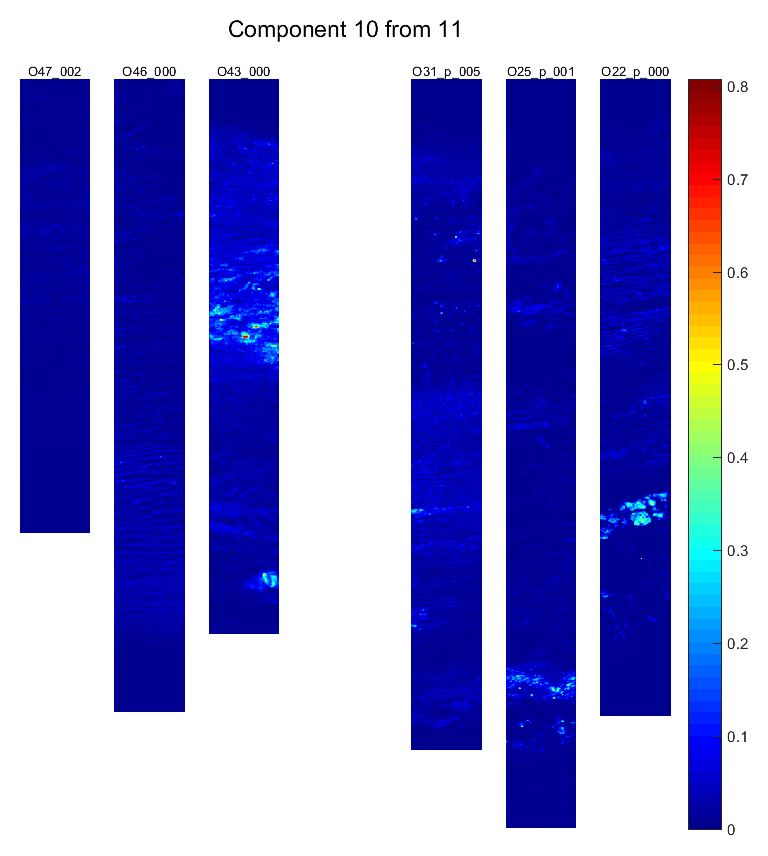}
\Capttwo{10}
\end{figure*}

\begin{figure*}[tb]
\includegraphics[width=0.38\linewidth]{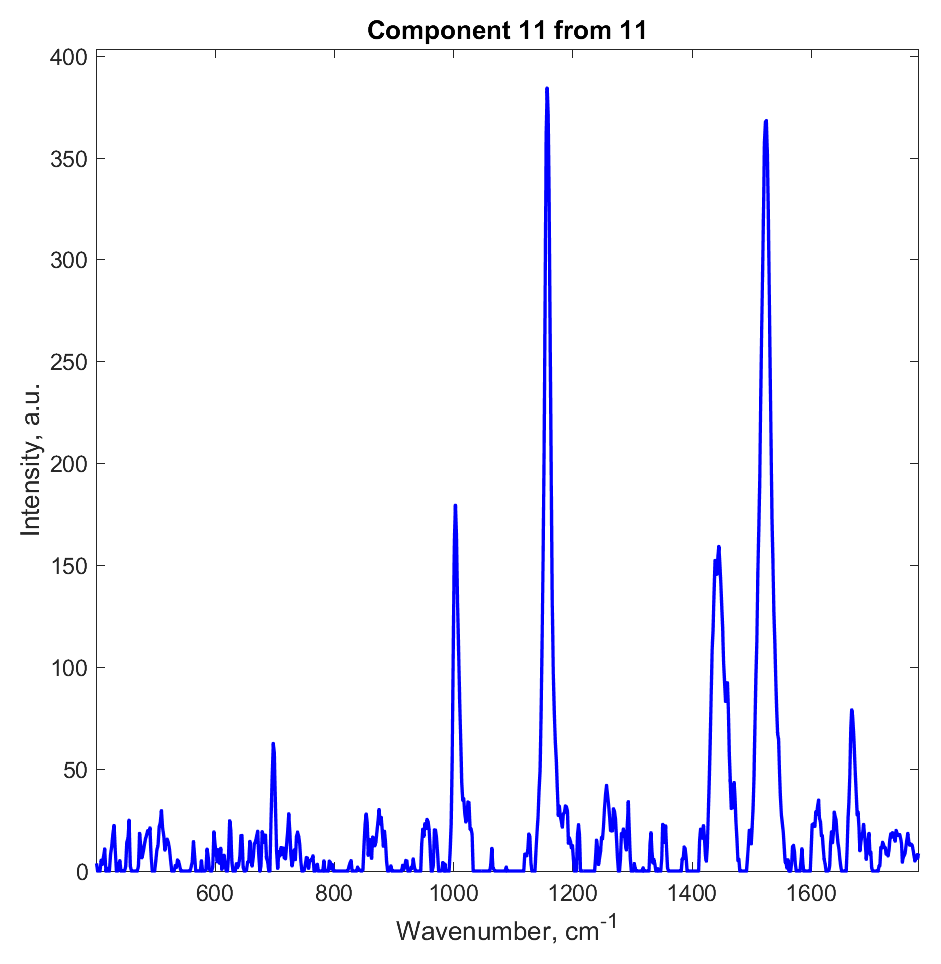}
\includegraphics[width=0.50\linewidth]{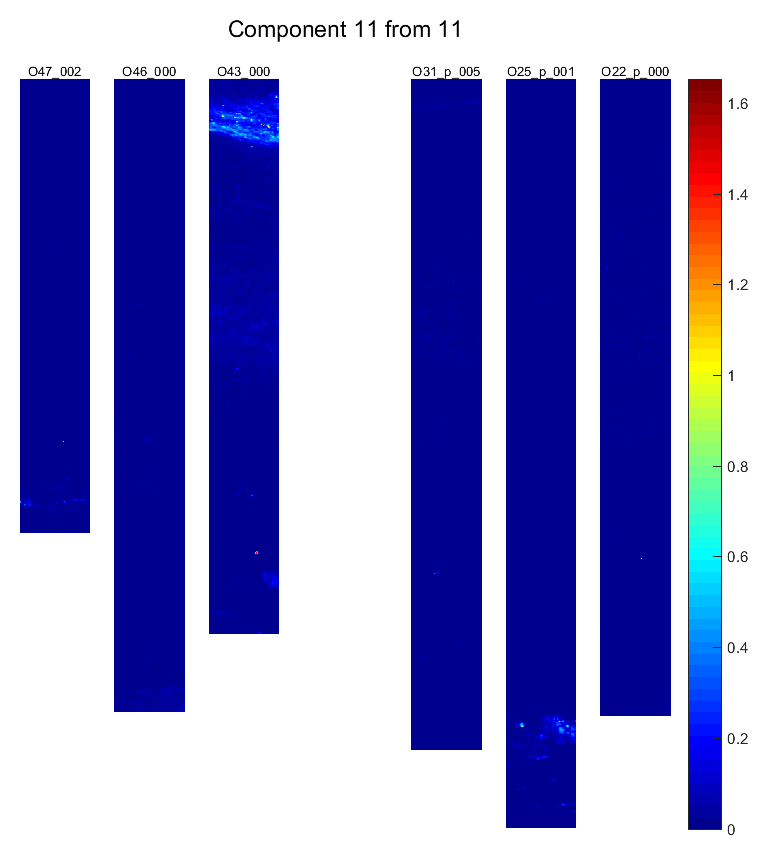}
\Capttwo{11}\label{Fig:Comp11}
\end{figure*}

\begin{figure*}[tb]
\includegraphics[width=0.38\linewidth]{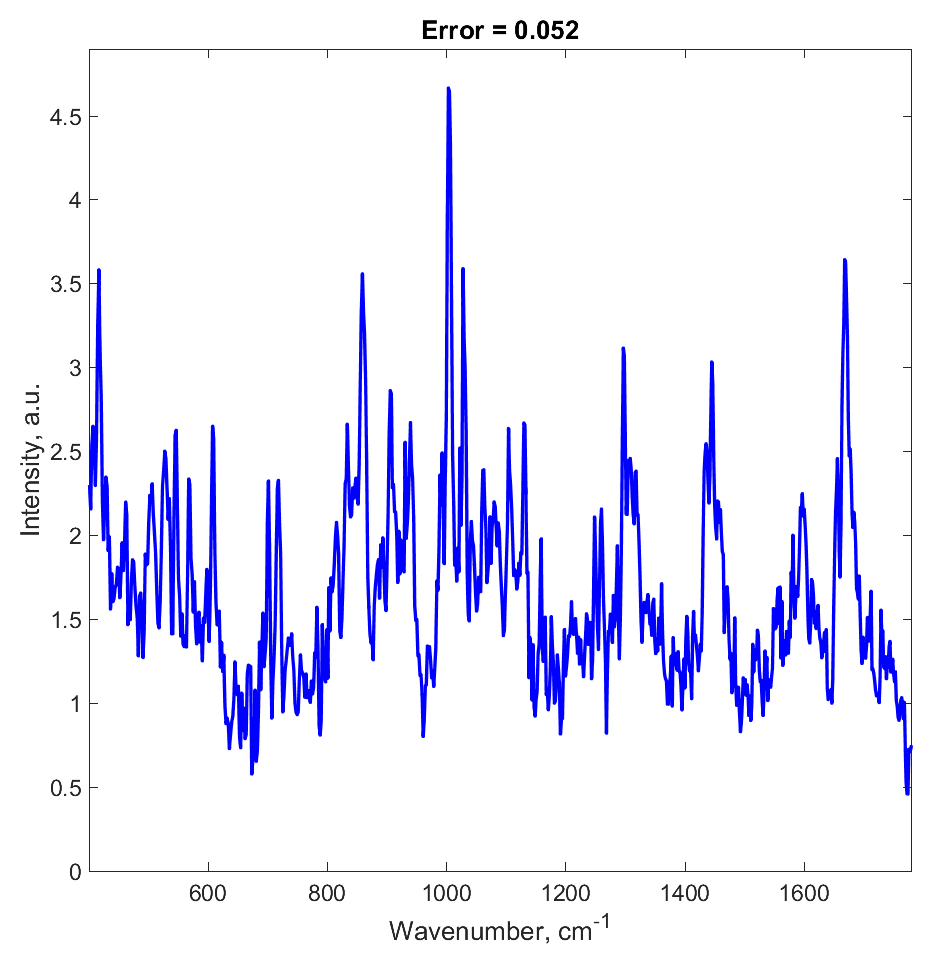}
\includegraphics[width=0.50\linewidth]{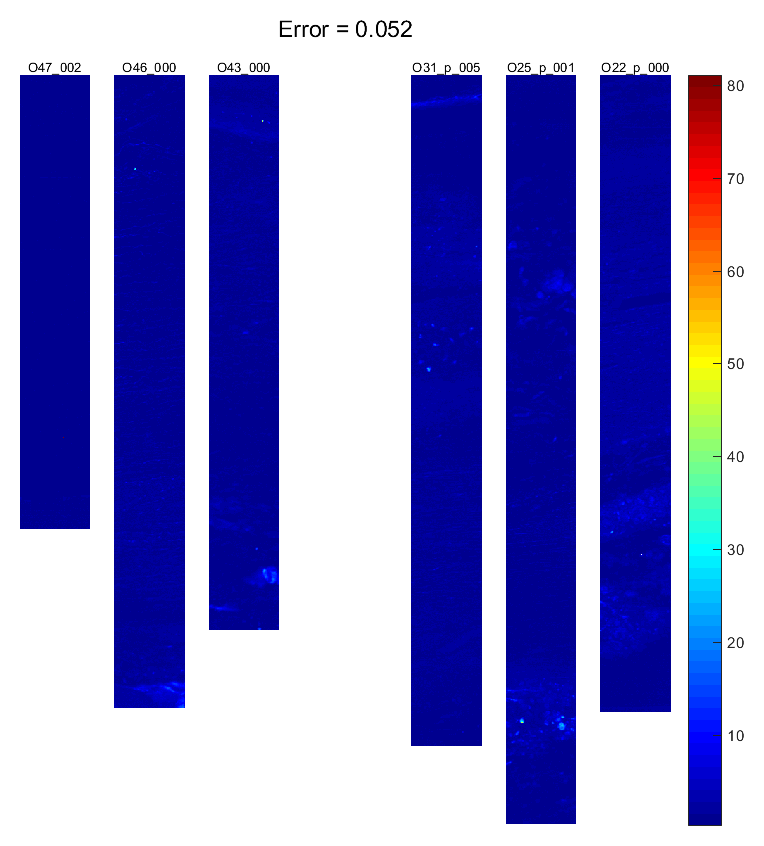}
\caption{Absolute factorization error averaged over image pixels~(left) and wavenumbers~(right). The relative factorization error $E(\C,\S)$ is 5.2\% as indicated in the title.}
\end{figure*}

\end{document}